\documentclass[12pt]{article}
\usepackage{graphicx}
\usepackage{amsmath}

\begin{document}

\hfill VPI-IPNAS-08-03

\hfill ILL-TH-08-2

\vspace{0.5in}

\begin{center}
{\large\bf A-twisted heterotic Landau-Ginzburg models}

\vspace{0.1in}

Josh Guffin$^1$, Eric Sharpe$^2$ \\

$\,$

\begin{tabular}{cc}
  \begin{tabular}{rl}
	 $^1$ \hspace*{-5.5mm} & Department of Physics\\
	                       & University of Illinois, Urbana-Champaign \\
	                       & 1110 West Green Street \\
	                       & Urbana, IL  61801-3080\\
  \end{tabular} &
  \begin{tabular}{rl}
	 $^2$ \hspace*{-5mm} & Physics Department\\
	                     & Robeson Hall (0435) \\
	                     & Virginia Tech\\
	                     & Blacksburg, VA  24061\\
  \end{tabular}
\end{tabular}

{\tt guffin@uiuc.edu}, 
{\tt ersharpe@vt.edu} \\

$\,$

\end{center}

In this paper, we apply the methods developed in recent work for constructing 
A-twisted (2,2) Landau-Ginzburg models to analogous (0,2) models.  In
particular, we study (0,2) Landau-Ginzburg models on topologically
non-trivial spaces away from large-radius limits, where one expects
to find correlation function contributions akin to (2,2) curve corrections.
Such heterotic theories admit A- and B-model twists, and exhibit a duality
that simultaneously exchanges the twists and dualizes the gauge bundle. 
We explore how this duality operates in heterotic
Landau-Ginzburg models, as well as other properties of these theories,
using examples which RG flow to heterotic nonlinear sigma models as checks
on our methods.

\begin{flushleft}
January 2008
\end{flushleft}

\newpage

\tableofcontents

\newpage

\section{Introduction}

Recently, there has been some significant progress in understanding
nonperturbative corrections in heterotic (or (0,2)) nonlinear sigma models,
see for example 
\cite{abs,ks1,ade,ksa,bchir,ksb,distshar,mason-skinner,ilarion1,ilarion2}.
In particular, (0,2) generalizations of the A and B model topological
field theories have been worked out, as well as (0,2) analogues of
quantum cohomology computations.

In our recent paper \cite{gs1}, we discussed the A twist of (2,2)
Landau-Ginzburg models, which has only rarely been discussed in the
literature before.  In this paper, we extend the analysis of
\cite{gs1} to consider 
(pseudo-)topological A, B twists of (0,2) Landau-Ginzburg
models, closely analogous to the A, B (0,2) twists of
nonlinear sigma models (NLSMs) discussed in
\cite{ks1,ade,ksa,bchir,ksb,ilarion1,ilarion2}.  

The A and
B twists of heterotic NLSMs are related by a duality that exchanges the
gauge bundle ${\cal E}$ with its dual ${\cal E}^{\vee}$.  In the (0,2)
case, the same duality exists, and also exchanges the two sets of
contributions to bosonic potentials and Yukawa couplings.  We discuss
anomaly cancellation, chiral rings, and correlation function computations
in (0,2) theories, and check our methods using examples of (0,2)
Landau-Ginzburg models that flow in the IR to heterotic NLSMs on nontrivial
spaces.

In section~\ref{sec:review}, we review some facts about (2,2)
Landau-Ginzburg models, including both A and B twists.  In particular, we
give some details of computations carried out in \cite{gs1} for the
A-twisted Landau-Ginzburg model on the total space of ${\cal
O}(-5)\rightarrow {\bf P}^4$,
which is in the same universality class as the ordinary A model on the
quintic hypersurface in ${\bf P}^4$.

In section~\ref{hetlggenl} we give an overview of heterotic Landau-Ginzburg 
models,
their A and B twists,
and dualities possessed by heterotic theories.  

In section~\ref{model:cokernel} we discuss heterotic Landau-Ginzburg models
in the same universality class as heterotic NLSMs whose gauge bundles
are described as cokernels of injective maps.  We discuss general aspects
and consistency checks of A and B twists, spectators, and discuss
two examples involving gauge bundles that are deformations of the tangent
bundle.

In section~\ref{hetb} we discuss heterotic Landau-Ginzburg models
in the same universality class as heterotic NLSMs whose gauge bundles
are described as kernels of surjective maps.

In section~\ref{sec:monad} we discuss the more general case of heterotic
Landau-Ginzburg models in the same universality class as heterotic NLSMs
whose gauge bundles are the cohomologies of short complexes, also known
as monads.

In section~\ref{sec:monadCI} we discuss the even more general case
of heterotic Landau-Ginzburg models in the same universality class as
heterotic NLSMs on complete intersections, whose gauge bundles are the 
cohomologies of monads.

Finally in section~\ref{app:hyper} we have an appendix deriving
a useful hypercohomology computation, that is often applied to chiral
ring computations.

\section{Review of (2,2) Landau-Ginzburg models}
\label{sec:review}

Here, we reiterate some key results of \cite{gs1}.
We begin by defining a (2,2) Landau-Ginzburg model as a NLSM together with a 
superpotential -- such a model requires that 
one must specify both a complex Riemannian manifold and a
superpotential (a holomorphic function over that manifold).

The most general Landau-Ginzburg model that one can write
down has the following action:
\begin{equation}
\begin{split}
\frac{1}{\alpha'} \left. \int_{\Sigma} d^2z \right( &
	 g_{\mu \nu} \partial \phi^{\mu} \overline{\partial} \phi^{\nu}
	  +  i B_{\mu \nu} \partial \phi^{\mu} \overline{\partial} \phi^{\nu}
	  + \frac{i}{2} g_{\mu \nu} \psi_-^{\mu} D_z \psi_-^{\nu} 
	  + \frac{i}{2} g_{\mu \nu} \psi_+^{\mu} D_{\overline{z}} \psi_+^{\nu} \\
	 & \left. 
	 + R_{i \overline{\jmath} k \overline{l}} \psi_+^i \psi_+^{\overline{\jmath}} \psi_-^k \psi_-^{\overline{l}}
	 + 2 g^{i \overline{\jmath}} \partial_i W \partial_{\overline{\jmath}} \overline{W} 
	 + \psi_+^i \psi_-^j D_i \partial_j W 
	 + \psi_+^{\overline{\imath}} \psi_-^{\overline{\jmath}} D_{\overline{\imath}}
	 \partial_{\overline{\jmath}} \overline{W} \vphantom \int \right),
  \end{split}
	\label{eq:22landauGinzburgAction}
\end{equation}
where $W$ is  the superpotential and
\begin{displaymath}
D_i \partial_j W \: = \: \partial_i \partial_j W \: - \:
\Gamma_{ij}^k \partial_k W.
\end{displaymath}
The fermions couple to the bundles
\begin{align*}
\psi_+^i \: &\in \: \Gamma_{ C^{\infty} }\left( K_{\Sigma}^{1/2} \otimes
\phi^* T^{1,0}X \right)                                         & \psi_-^i
\: &\in \: \Gamma_{ C^{\infty} }\left( \overline{K}_{\Sigma}^{1/2} \otimes \left( \phi^* T^{0,1} X\right)^{\vee} \right) \\
\psi_+^{\overline{\imath}} \: &\in \: \Gamma_{ C^{\infty} }\left(
K_{\Sigma}^{1/2} \otimes \left( \phi^* T^{1,0}X \right)^{\vee} \right) &
\psi_-^{\overline{\imath}} \: &\in \: \Gamma_{ C^{\infty} } \left(
\overline{K}_{\Sigma}^{1/2} \otimes \phi^* T^{0,1} X \right),
\end{align*}
where $K_{\Sigma}$ denotes the canonical bundle on the worldsheet
$\Sigma$.
The bosonic potential is of the form $\sum_i | \partial_i W |^2$.
Even though such models are not even classically scale-invariant, we can
use them to define conformal field theories by applying renormalization
group  flow -- the endpoint of which is a (possibly trivial) conformal field theory.

\subsection{B-twisted models}
\label{ssec:22BLG}

The B twist of a Landau-Ginzburg model is defined by taking the
fermions to be the sections of the bundles
\begin{align*}
\psi_+^i \: &\in \: \Gamma_{ C^{\infty} }\left( K_{\Sigma} \otimes \phi^*
T^{1,0}X \right)                      & \psi_-^i \: &\in \: \Gamma_{ C^{\infty} }\left( \overline{K}_{\Sigma} \otimes \left( \phi^* T^{0,1} X\right)^{\vee} \right) \\
\psi_+^{\overline{\imath}} \: &\in \: \Gamma_{ C^{\infty} }\left( \left(
\phi^* T^{1,0}X \right)^{\vee} \right) & \psi_-^{\overline{\imath}} \: &\in
\: \Gamma_{ C^{\infty} } \left( \phi^* T^{0,1} X \right)
\end{align*}
and combining them as
\begin{align*}
\eta^{\overline{\imath}} & = \psi_+^{\overline{\imath}} \: + \:
\psi_-^{\overline{\imath}} &
\rho_z^i & = \psi_+^i \\
\theta_i & = g_{i \overline{\jmath}} \left( \psi_+^{\overline{\jmath}}
\: - \: \psi_-^{\overline{\jmath}} \right)  &
\rho_{\overline{z}}^i & = \psi_-^i.
\end{align*}
The operators in the chiral ring for this theory consist of 
BRST-closed (mod BRST-exact) products of the form
\begin{displaymath}
b(\phi)^{j_1 \cdots j_m}_{\overline{\imath}_1 \cdots \overline{\imath}_n}
\eta^{\overline{\imath}_1} \cdots \eta^{\overline{\imath}_n}
\theta_{j_1} \cdots \theta_{j_m}.
\end{displaymath}
When the superpotential is not constant, $\theta$ is no longer
BRST-invariant and such operators may not be interpreted as sheaf
cohomology of exterior powers of the tangent bundle. Rather they comprise
the hypercohomology of the complex
\begin{equation} 
\ldots \: \stackrel{i_{dW}}{\longrightarrow} \: \Lambda^{3}TX \:
\stackrel{i_{dW}}{\longrightarrow} \:
\Lambda^{2}TX \: \stackrel{i_{dW}}{\longrightarrow}  \:
TX \: \stackrel{i_{dW}}{\longrightarrow} \:
{\cal O}_{X},
\label{bmodcpx}
\end{equation}
as discussed in \cite{gs1}.

\subsection{A-twisted models}
\label{ssec:22ALG}

Unlike the B-twisted model, we must make special adjustments to the
A-twisted model to take into account the fact that the Yukawa couplings 
\begin{displaymath}
 \psi_+^i \psi_-^j D_i \partial_j W \: + \:
 \psi_+^{\overline{\imath}} \psi_-^{\overline{\jmath}} D_{\overline{\imath}}
\partial_{\overline{\jmath}} \overline{W} 
\end{displaymath}
arising from a non-trivial superpotential become one-forms on the
worldsheet with the na\"ive twisting.  One approach to solving this problem
involves further twisting a subset of the chiral superfields by a $U(1)$
isometry in such a way that the terms become either functions or two-forms
on the worldsheet.  In the special case of total spaces of holomorphic vector
bundles on complex manifolds with superpotential given by a section of the
dual bundle contracted with the fiber coordinates, one twists
the the fiber coordinates 
\cite{gs1}.

It will be useful for discussions of (0,2) models to mention an example
from in \cite{gs1}, where correlation functions were computed in a theory
in the same universality class as the A model on the quintic hypersurface
in ${\bf P}^4$.  Specifically, this example was a Landau-Ginzburg
model over the total space of ${\cal O}(-5) \rightarrow {\bf P}^4$,
with superpotential $W = pG$, $p$ a fiber coordinate on ${\cal O}(-5)$, $G$
a section of ${\cal O}(5)$ whose vanishing locus is the quintic
in question.
In this example, the fiber coordinate $p$ 
was twisted so that the fermions in the theory coupled to bundles as
\begin{align*}
\psi_+^i ( \equiv \chi^i )                                        & \in  \Gamma_{C^{\infty}}\left( \phi^* T^{1,0}{\bf P}^4 \right)                             & \psi_-^i ( \equiv \psi_{\overline{z}}^i )                            & \in  \Gamma_{C^{\infty}}\left( \overline{K}_{\Sigma} \otimes (\phi^* T^{0,1}{\bf P}^4)^{\vee} \right) \\
\psi_+^{\overline{\imath}} ( \equiv \psi_z^{\overline{\imath}} )  & \in \Gamma_{C^{\infty}}\left( K_{\Sigma} \otimes (\phi^* T^{1,0}{\bf P}^4)^{\vee} \right) & \psi_-^{\overline{\imath}} ( \equiv \chi^{\overline{\imath}} )       & \in  \Gamma_{C^{\infty}}\left( \phi^* T^{0,1} {\bf P}^4 \right).\\
\psi_+^p (\equiv \psi_z^p)                                        & \in  \Gamma_{C^{\infty}}\left( K_{\Sigma} \otimes \phi^* T^{1,0}_{\pi} \right)             & \psi_-^p (\equiv \chi^p)                                             & \in  \Gamma_{C^{\infty}}\left( ( \phi^* T^{0,1}_{\pi} )^{\vee} \right) \\
\psi_+^{\overline{p}} ( \equiv \chi^{\overline{p}} )              & \in  \Gamma_{C^{\infty}}\left( ( \phi^* T^{1,0}_{\pi} )^{\vee} \right)                     & \psi_-^{\overline{p}} ( \equiv \psi_{\overline{z}}^{\overline{p}} )  & \in  \Gamma_{C^{\infty}}\left( \overline{K}_{\Sigma}\otimes \phi^* T^{0,1}_{\pi} \right).
\end{align*}

Topological operators then correspond to d-closed differential forms on ${\bf 
P}^4$,
\begin{eqnarray*}
\lefteqn{
b_{i_1 \cdots i_n \overline{\jmath}_1 \cdots
{\jmath}_m}(\phi) \chi^{i_1} \cdots \chi^{i_n}
\chi^{\overline{\jmath}_1} \cdots
\chi^{\overline{\jmath}_m} 
} \\
& \hspace*{1in}  \leftrightarrow &
b_{i_1 \cdots i_n \overline{\jmath}_1 \cdots
{\jmath}_m}(\phi)
dz^{i_1} \wedge \cdots \wedge dz^{i_n} \wedge
d \overline{z}^{\overline{\jmath}_1} \wedge \cdots
\wedge
d \overline{z}^{\overline{\jmath}_m}.
\end{eqnarray*}
under the further restriction that evaluating the path integral causes
factors to be brought down in such a way that the forms are to be
interpreted as forms on the zero locus of the section defining the
superpotential -- in this case, the quintic.

Classical contributions to correlation functions were analyzed.  Since the
superpotential terms are BRST-exact, the superpotential may be rescaled
without affecting correlation functions.  For example, the classical
contribution to a genus-zero correlation function $\langle {\cal O}_1
\cdots {\cal O}_n \rangle$ as studied in section $3.3.1$ of
\cite{gs1} is 
\begin{equation}
\begin{split}
\langle {\cal O}_1 \cdots {\cal O}_n \rangle &= \int_{ {\bf P}^4 } d^2 \phi^i \int \prod_i d \chi^i d \chi^{ \overline{\imath}} d \chi^p d \chi^{\overline{p}} {\cal O}_1 \cdots {\cal O}_n \\
& \hspace*{1cm} \cdot \exp\left(
- 2 g^{p \overline{p}} | \lambda G(\phi) |^2
- \lambda \chi^i \chi^p D_i G
- \bar \lambda \chi^{\overline{p}} \chi^{\overline{\jmath}} D_{\jmath} \overline{G}
- R_{i \overline{p} p \overline{k} } \chi^i \chi^{\overline{p}} \chi^p \chi^{\overline{k}} \right).
\label{eq:quinticMQ}
\end{split}
\end{equation}
One may examine such a correlator in the $\lambda \rightarrow \infty$ limit,
wherein
\begin{align}
\langle {\cal O}_1 \cdots {\cal O}_n \rangle
&= \int_{ {\bf P}^4 } d^2 \phi^i \int d \chi \; {\cal O}_1 \cdots {\cal O}_n \;
\vert \chi^i D_i G \vert ^2 \exp \left( - 2 g^{p \overline{p}} | G(\phi) |^2 \right)  
\label{eq:quinticLimit0}
\intertext{and the $\lambda \rightarrow 0$ limit, to wit}
\langle {\cal O}_1 \cdots {\cal O}_n \rangle &= \int_{ {\bf P}^4 } d^2
\phi^i \int d \chi \; {\cal O}_1 \cdots {\cal O}_n \;
R_{i \overline{p} p \overline{k} } g^{p\overline p} \chi^i 
\chi^{\overline{k}}.
\label{eq:quinticLimitInf}
\end{align}
Here, $d\chi$ stands for an integral over each of the $\chi^i$ 
zero modes, as well as those of their complex conjugates, while the $\chi^p$ 
zero modes and their conjugates have been integrated out.

The $d\chi^p$ integral of the exponential in \eqref{eq:quinticMQ} is
interpreted as a Mathai-Quillen form of the vertical subbundle in
$T{\cal O}(5)$, pulled back to ${\bf P}^4$ by the section $G$.
The two limits are respectively as an integral over the normal bundle to
the quintic in ${\bf P}^4$ with insertions of factors normal to quintic,
and as an integral over ${\bf P}^4$ with an insertion of the Euler class of 
${\cal O}(5)$.
We shall consider a heterotic model deforming this theory in
section~\ref{ex:quinticdef}.

\section{General aspects of (0,2) Landau-Ginzburg models}

\label{hetlggenl}

\subsection{Untwisted models}

Physically, a (0,2) Landau-Ginzburg model is described by a collection of
``(0,2) chiral'' superfields $\Phi^i = (\phi^i, \psi_+^i)$,
corresponding to local coordinates on a complex K\"ahler manifold $X$, 
and ``(0,2) fermi'' superfields
$\Lambda^{\alpha} = (\lambda_-^{\alpha}, H^{\alpha})$, which transform
as local smooth sections of some holomorphic
vector bundle ${\cal E}$ over $X$.
Heterotic Landau-Ginzburg models are also NLSMs with
superpotentials, though here the NLSM possesses only
(0,2) supersymmetry, and the superpotential is a Grassmann-odd holomorphic
function of the superfields, rather than an ordinary holomorphic function.
For a superpotential of the form 
\begin{equation}
W \: = \: \Lambda^{\alpha} F_{\alpha}(\Phi),
\label{eq:superpotentialForm}
\end{equation}
where $F_a \in \Gamma( X, {\cal E}^{\vee} )$,
and in which, in superspace notation, $\overline{D}_+ \Lambda^a = E^a(\phi)$
for $E^a \in \Gamma(X, {\cal E})$,
the action for a heterotic
Landau-Ginzburg  model has the form
\begin{equation}
\begin{split}
\left. \frac{1}{\alpha'} \int_{\Sigma} d^2z \right(
\left (g_{\mu \nu} +  i B_{\mu \nu} \right) \partial \phi^{\mu} \overline{\partial} \phi^{\nu} 
&+ \frac{i}{2} g_{\mu \nu} \psi_+^{\mu} D_{\overline{z}} \psi_+^{\nu}
+ \frac{i}{2}h_{\alpha \beta} \lambda_-^{\alpha} D_{z} \lambda_-^{\beta}
+  F_{i \overline{\jmath} a \overline{b}} \psi_+^i \psi_+^{\overline{\jmath}} \lambda_-^a \lambda_-^{\overline{b}} \\
& 
+  2 h^{a \overline{b}} F_a \overline{F}_{\overline{b}}
+  \psi_+^i \lambda_-^a D_i F_a 
+ \psi_+^{\overline{\imath}} \lambda_-^{\overline{b}} D_{\overline{\imath}} \overline{F}_{\overline{b}} \\
& \left.
+  2 h_{a \overline{b}} E^a \overline{E}^{\overline{b}} 
+ \psi_+^i \lambda_-^{\overline{a}} \left( D_i E^b \right) h_{\overline{a} b}
+ \psi_+^{\overline{\imath}} \lambda_-^a \left( D_{\overline{\imath}} \overline{E}^{\overline{b}} \right) h_{a \overline{b} } \vphantom \int \right).
\label{eq:02LGAction}
\end{split}
\end{equation}
In this expression, $h_{\alpha \beta}$ is a Hermitian fiber metric on ${\cal E}$,
and $A$ is a connection on ${\cal E}$ determined by the hermitian fiber
metric $h$ as\footnote{When the fiber metric $h$ matches 
the metric $g$ on the $(2,2)$ locus, $A$ matches the Christoffel connection.}
\begin{align*}
A_{\mu  b}^a                          & =  h^{a \overline{c}} \partial_{\mu} h_{\overline{c} b}\\
A_{\mu   \overline{b}}^{\overline{a}} & = h^{\overline{a} b} \partial_{\mu} h_{b \overline{c}}\;.
\end{align*}
The covariant derivatives $D \lambda_-$ are given by
\begin{displaymath}
D_z \lambda_-^{\alpha} \: = \:
\partial \lambda_-^{\alpha} \: + \: \left( \partial \phi^{\mu} \right)
A_{\mu \: \: \beta}^{\alpha} \lambda_-^{\beta}.
\end{displaymath}
More formally, the fermions couple to bundles as follows:
\begin{align*}
\psi_+^i  &\in  \Gamma_{C^{\infty}}\left( K_{\Sigma}^{1/2} \otimes \phi^* T^{1,0}X \right)                                            & \lambda_-^a  &\in  \Gamma_{ C^{\infty} }\left( \overline{K}_{\Sigma}^{1/2} \otimes \left( \phi^* \overline{\cal E} \right)^{\vee} \right) \\
\psi_+^{\overline{\imath}}  &\in  \Gamma_{ C^{\infty} }\left( K_{\Sigma}^{1/2} \otimes \left( \phi^* T^{1,0} X \right)^{\vee} \right) & \lambda_-^{\overline{a}}  &\in  \Gamma_{ C^{\infty} }\left( \overline{K}_{\Sigma}^{1/2} \otimes \phi^* \overline{ {\cal E} } \right).
\end{align*}
There is a constraint on $E^a$ and $F_a$, namely that
\begin{displaymath}
\sum_a E^a(\phi) F_a(\phi) \: = \: 0.
\end{displaymath}
More general superpotentials are possible in principle --
we could consider $\Lambda^a \Lambda^b \Lambda^c F_{abc}(\phi)$,
which would be subject to the constraint
\begin{displaymath}
E^a(\phi) \Lambda^b \Lambda^c F_{abc}(\phi) \: + \:
\Lambda^a E^b(\phi) \Lambda^c F_{abc}(\phi) \: + \:
\Lambda^a \Lambda^b E^c(\phi) F_{abc}(\phi) \: = \: 0,
\end{displaymath}
or more simply
\begin{displaymath}
E^a(\phi)\left( F_{abc}(\phi) \: + \: F_{bac}(\phi)
 \: + \: F_{bca}(\phi) \right)
\: = \: 0.
\end{displaymath}
We shall not treat such forms, as the expression
\eqref{eq:superpotentialForm} suffices to describe everything currently in
the literature.

An ordinary heterotic NLSM, where each
$E^a$ and $F_a$ vanishes, is only well-defined in the special case
that $c_2({\cal E}) = c_2(TX)$.  The same condition is applies here,
just as the B-twisted Landau-Ginzburg model is only well-defined
over spaces obeying the same consistency condition as the
underlying B-twisted NLSM.
Thus, we will only consider heterotic Landau-Ginzburg models over
spaces $X$ with holomorphic vector bundles ${\cal E}$ such that
$c_2({\cal E}) = c_2(TX)$.

The supersymmetry transformations in this model have the following form:
\begin{alignat*}{3}
   & \delta \phi^i                     & & ={} & & i \alpha_- \psi_+^i \\
   & \delta \phi^{\overline{\imath}}   & & =   & & i \tilde{\alpha}_- \psi_+^{ \overline{\imath}} \\
   & \delta \psi_+^i                   & & =   & & - \tilde{\alpha}_- \partial \phi^i \\
   & \delta \psi_+^{\overline{\imath}} & & =   & & - \alpha_- \partial \phi^{\overline{\imath}} \\
   & \delta \lambda_-^a                & & =   & & - i \alpha_- \psi_+^j A_{j \: \: c}^a \lambda_-^c \: + \: i \alpha_- h^{a \overline{b}} \overline{F}_{\overline{b}} \: + \: i \tilde{\alpha}_- E^a\\
   & \delta \lambda_-^{\overline{a}}   & & =   & & - i \tilde{\alpha}_- \psi_+^{\overline{\jmath}}  A_{\overline{\jmath} \: \: \overline{c}}^{\overline{a}} \lambda_-^{\overline{c}} \: + \: i \tilde{\alpha}_- h^{\overline{a} b} F_b \: + \: i \alpha_- \overline{E}^{\overline{a}}.
\end{alignat*}

One may recover the (2,2) Landau-Ginzburg model defined in
\eqref{eq:22landauGinzburgAction} as a special case by 
\begin{enumerate}
\item Assigning a corresponding (0,2) Fermi superfield $\Lambda^i$
to each (0,2) chiral superfield $\Phi^i$, so that ${\cal E} = TX$
\item Taking all $E^i \equiv 0$ 
\item Defining the (0,2) superpotential to be $\Lambda^i \partial_i
W (\Phi)$, with $W$ is the corresponding (2,2) superpotential.
\end{enumerate}

Note that the Lagrangian \eqref{eq:02LGAction}
is symmetric under the exchange
${\cal E} \leftrightarrow {\cal E}^{\vee}$
(which exchanges $\lambda_-^a$ with 
$\lambda_{- a} = h_{a \overline{b}} \lambda_-^{\overline{b}}$),
so long as we simultaneously exchange $E^a(\phi) \leftrightarrow
F_a(\phi)$.  This duality was discussed in the context of
the (0,2) A, B model analogues in \cite{ade,bchir}.

\subsection{Topological twists}
\label{ssec:topTwists}

Next, let us discuss the (pseudo-)topological twists of this theory.

Following \cite{bchir}, we can define an analogue of the B-twisting for
heterotic theories.  In the case $E^a \equiv 0$ (though $F_a$ can be
nonzero), we define the twisting by taking the fermions to couple to the
following bundles:
\begin{align*}
\psi_+^i \: &\in \: \Gamma_{C^{\infty}}\left( K_{\Sigma} \otimes \phi^* T^{1,0}X \right)                         & \lambda_-^a \: &\in \: \Gamma_{ C^{\infty} }\left( \overline{K}_{\Sigma} \otimes \left( \phi^* \overline{\cal E} \right)^{\vee} \right) \\
\psi_+^{\overline{\imath}} \: &\in \: \Gamma_{ C^{\infty} }\left( \left( \phi^* T^{1,0} X \right)^{\vee} \right) & \lambda_-^{\overline{a}} \: &\in \: \Gamma_{ C^{\infty} }\left( \phi^* \overline{ {\cal E} } \right).
\end{align*}
Indeed, as a consistency check,
recall that the (0,2) Landau-Ginzburg model reduces to the (2,2)
Landau-Ginzburg model when $E^a \equiv 0$ (among other things),
and furthermore the B-twist of the (2,2) Landau-Ginzburg model
can be defined in the same way as for the NLSM;
no R-symmetry is required.
When the $E^a \not\equiv 0$, on the other hand, to perform the topological
twist we will have to proceed in a fashion analogous to that we have
discussed for A-twisted Landau-Ginzburg models.

In \cite{bchir} it was observed that the (0,2) B-twisted heterotic
NLSM was only well-defined when $\Lambda^{top}
{\cal E} \cong K_X$, in addition to the usual constraint that
$\mbox{ch}_2({\cal E}) = \mbox{ch}_2(TX)$.  This new condition arose
in order to make the path integral measure well-defined.
The same condition arises in the B-twisted heterotic Landau-Ginzburg
model discussed here, independent of the superpotential defined by the
$F_a$ (so long as all $E^a \equiv 0$, so that the na\"ive B twisting
is sensible).

Furthermore, when all the $E^a \equiv 0$, the chiral ring of the
(0,2) B twist is straightforward to compute.
Just as in the ordinary (2,2) B model, where one can create a BRST-invariant
fermion $\theta_i$ by lowering $\psi_{\pm}^{\overline{\imath}}$
indices with the metric, note here that the BRST variation of
$h_{a \overline{b}} \lambda_-^{\overline{b}}$ is given by
\begin{displaymath}
\delta \left( h_{ a \overline{b}} \lambda_-^{\overline{b}} \right)
\: = \: + i  \tilde{\alpha}_- F_a.
\end{displaymath}
When all the $F_a \equiv 0$ in addition to the $E^a \equiv 0$, the
$\lambda_{- a}$'s are BRST-invariant and we recover the standard result
\cite{bchir,dg} that the part\footnote{ The entire chiral ring is larger,
taking into account for example the fact that the BRST variation $\delta
\lambda_-^a = 0$.  However, just as in \cite{ks1,bchir}, we ignore those
more general elements, as we neither have nor expect a clean description of
their correlation functions.  } of the chiral ring generalizing the (2,2)
case is built from products of $\lambda_{- a} = h_{a \overline{b}}
\lambda_-^{\overline{b}}$, and is described by sheaf cohomology
$H^{\cdot}(X, \Lambda^{\cdot} {\cal E})$.  When the $F_a$ do not all
vanish, then just as for B-twisted (2,2) Landau-Ginzburg models in
section~\ref{ssec:22BLG}, we get immediately that the relevant part of the
chiral ring of the (0,2) B-twisted Landau-Ginzburg model should be given by
hypercohomology of the complex
\begin{displaymath}
\cdots \: \stackrel{ i_{F_a} }{\longrightarrow}
 \: \Lambda^2 {\cal E} \: \stackrel{ i_{F_a} }{\longrightarrow} \: 
{\cal E} \: \stackrel{ i_{F_a} }{\longrightarrow} \:
{\cal O}_X.
\end{displaymath}
Note that this correctly specializes not only to the (0,2) B model 
ring $H^{\cdot}(X, \Lambda^{\cdot} {\cal E})$ when the $F_a \equiv 0$, 
but also to the (2,2)
B-twisted Landau-Ginzburg model chiral ring.
This is because the (2,2) B model in (0,2) language is defined by
${\cal E} = TX$ and $F_i = \partial_i W$, so that the complex above
reduces to \eqref{bmodcpx}, whose hypercohomology
is the chiral ring of the (2,2) B-twisted Landau-Ginzburg model.

We can also define an analogue of the 
A-twisting for heterotic theories.
In the case $F_a \equiv 0$ (but $E^a$ can be nonzero), 
we twist by taking the fermions to couple to
the following bundles:
\begin{align*}
\psi_+^i \: (\equiv \chi^i) \: &\in \: \Gamma_{C^{\infty}}\left( \phi^* T^{1,0}X \right)                                                                                   & \lambda_-^a \: &\in \: \Gamma_{ C^{\infty} }\left( \overline{K}_{\Sigma} \otimes \left( \phi^* \overline{\cal E} \right)^{\vee} \right) \\
\psi_+^{\overline{\imath}} \: (\equiv \psi_z^{\overline{\imath}}) \: &\in \: \Gamma_{ C^{\infty} }\left( K_{\Sigma} \otimes \left( \phi^* T^{1,0} X \right)^{\vee} \right) & \lambda_-^{\overline{a}} \: &\in \: \Gamma_{ C^{\infty} }\left( \phi^* \overline{ {\cal E} } \right).
\end{align*}
On the other hand, when the $F_a \not\equiv 0$,
then to make sense of the A twist we have to work harder, using 
a nontrivial R-symmetry.
Note that this is symmetric with the behavior of the
(0,2) B model analogue.  Indeed, as discussed earlier,
there is a symmetry under which 
${\cal E} \leftrightarrow {\cal E}^{\vee}$ and $E^a \leftrightarrow
F_a$, which for cases without superpotential exchanges the
(0,2) A and B analogues \cite{ade,bchir}.
Since such a symmetry exists for the theories without superpotential,
one might hope that it would also exist in cases with a superpotential,
and that is exactly what we have observed.

In \cite{ks1,bchir} it was shown that the path integral measure in a 
(0,2) A-twisting of a heterotic
NLSM is only well-defined when
$\Lambda^{top} {\cal E}^{\vee} \cong K_X$, in addition to the usual
anomaly-cancellation constraint $\mbox{ch}_2({\cal E}) = 
\mbox{ch}_2(TX)$. In the present case, for a theory with nonvanishing
$E^a$ (but all $F_a \equiv 0$), it is straightforward to see that the
same constraint should be imposed.

Furthermore, just as in the (0,2) B twist above, when $F_a \equiv 0$
we can also get a clean result for the part of the chiral ring
generalizing that of (2,2) models.
Here in this (0,2) A twist, note that the BRST variation
$\delta \lambda_-^{\overline{a}} = i  \alpha_- \overline{E}^{
\overline{a}}$, and so just as for the (0,2) B twist,
the part of the chiral ring of the (0,2) A twist that generalizes
that of (2,2) models and of (0,2) A-twisted NLSMs
is given by the hypercohomology of the complex
\begin{displaymath}
\cdots \: \stackrel{ i_{E^a} }{\longrightarrow} \:
\Lambda^2 {\cal E}^{\vee} \: \stackrel{ i_{E^a} }{\longrightarrow} \:
{\cal E}^{\vee} \: \stackrel{ i_{E^a} }{\longrightarrow} \:
{\cal O}_X.
\end{displaymath}
In the case that both $E^a \equiv 0$ and $F_a \equiv 0$, 
this hypercohomology reduces to $H^{\cdot}(X, \Lambda^{\cdot} {\cal E}^{\vee})$,
exactly the result discussed in \cite{ks1,dg}.
Also as expected, 
the chiral ring of the (0,2) A-twisted
Landau-Ginzburg model is mapped into
the chiral ring for the dual (0,2) B-twisted Landau-Ginzburg model
by the duality discussed previously.

\section{Models realizing cokernels of maps}   \label{model:cokernel}

\subsection{General analysis of A twist}

Our first example will involve a heterotic Landau-Ginzburg model
that should flow under the renormalization group to a heterotic
NLSM on a space $B$ 
with a bundle ${\cal E}'$ defined as the
cokernel of an injective map:
\begin{displaymath}
{\cal E}' \: = \: \mbox{coker }\left\{
{\cal F}_1 \: \stackrel{\tilde{E}}{\longrightarrow} \: {\cal F}_2 \right\}.
\end{displaymath}
The corresponding heterotic Landau-Ginzburg model will be defined
over the space 
\begin{displaymath}
X \: = \: \mbox{Tot}\left( {\cal F}_1 \: \stackrel{\pi}{\longrightarrow}
\: B \right),
\end{displaymath}
with gauge bundle ${\cal E} \equiv \pi^* {\cal F}_2$,
all $F_a \equiv 0$,
and $E^a = p \tilde{E}^a$ for
$p$ fiber coordinates along ${\cal F}_1$.

In principle this is just a straightforward generalization of examples
in \cite{dk}, so knowledgeable readers will not find it surprising
that these two theories are related by renormalization group flow,
but let us take a moment to outline the arguments for completeness.
First, the potential term $h_{a \overline{b}} E^a 
\overline{E}^{\overline{b}}$ becomes
\[h_{a \overline{b}} |p|^2 \tilde{E}^a \overline{\tilde{E}}^{\overline{b}},\]
and since the $\tilde{E}$'s are assumed injective, this will act as a mass
term for the $p$ fields, so that the IR theory should see $B$ instead of
$X$.  Similarly, there is a Yukawa coupling term of the form
$i \psi_+^p \lambda_-^{\overline{a}} \tilde{E}^b h_{\overline{a} b}$,
which shows how the $\psi_+^p$ superpartner to the $p$ field pairs up with the
image of ${\cal F}_1$ to become massive, so that the remaining massless
$\lambda_-^a$ fermions see ${\cal E}'$.
This just gives an intuitive justification that the two theories
are related by renormalization group flow, but next, as part of
a discussion of general features of these models, we will see
that anomalies and chiral rings match, which give a more solid check. 

Before moving to a more specific example, let us first check some general
properties of the models we have just described.
For example, if anomaly cancellation holds in the heterotic NLSM on $B$ with gauge bundle ${\cal E}'$, then it had also
better hold in the heterotic Landau-Ginzburg model.
Anomaly cancellation in the NLSM on $B$ is the
statement that
\begin{displaymath}
\mbox{ch}_2(TB) \: = \: \mbox{ch}_2( {\cal E}' ) \: =  \:
\mbox{ch}_2({\cal F}_2) \: - \: \mbox{ch}_2({\cal F}_1).
\end{displaymath}
Anomaly cancellation in the heterotic Landau-Ginzburg model over $X$,
on the other hand, is the statement that
\begin{displaymath}
\mbox{ch}_2(TX) \: = \: \mbox{ch}_2({\cal E}) \: = \:
\pi^* \mbox{ch}_2({\cal F}_2).
\end{displaymath}
To relate $\mbox{ch}_2(TX)$ to the Chern classes of $TB$ and
${\cal F}_1$, we can use the short exact sequence
\begin{displaymath}
0 \: \longrightarrow \:
\pi^* {\cal F}_1 \: \longrightarrow \: TX \: \longrightarrow \:
\pi^* TB \: \longrightarrow \: 0,
\end{displaymath}
from which we read off that
\begin{displaymath}
\mbox{ch}_2(TX) \: = \: \pi^* \mbox{ch}_2 (TB) \: + \:
\pi^* \mbox{ch}_2({\cal F}_1),
\end{displaymath}
and hence the anomaly-cancellation conditions
in the NLSM and in the Landau-Ginzburg model
are equivalent:  anomaly cancellation holds in one
model if and only if it also holds in the other.

Since the $F_a \equiv 0$ in these heterotic Landau-Ginzburg models,
one should be able to perform the A twist na\"ively, without
twisting any bosonic fields.
When one does so, the consistency condition in the heterotic Landau-Ginzburg
model for the twist to be sensible is that
\begin{displaymath}
\Lambda^{top} {\cal E}^{\vee} \: \cong \: K_X.
\end{displaymath}
Comparing to the heterotic NLSM,
\begin{displaymath}
K_X \: \cong \: \pi^* \left( K_B \otimes \Lambda^{top} {\cal F}_1^{\vee}
\right), \: \: \:
\Lambda^{top} {\cal E}^{\vee} \: \cong \:
\pi^* \Lambda^{top} {\cal F}_2^{\vee},
\end{displaymath}
so we see that the condition for the A twist of the heterotic Landau-Ginzburg
model to be consistent is just
\begin{displaymath}
\pi^* \Lambda^{top} {\cal F}_2^{\vee} \: \cong \:
\pi^* K_B \otimes \pi^* \Lambda^{top} {\cal F}_1^{\vee},
\end{displaymath}
or equivalently
\begin{displaymath}
\pi^* K_B \: \cong \:
\pi^* \Lambda^{top} {\cal F}_1 \otimes
\pi^* \Lambda^{top} {\cal F}_2^{\vee} \: \cong \: \pi^* 
\Lambda^{top} {\cal E}'^{\vee},
\end{displaymath}
which is the pullback of the constraint that the A twist of the
corresponding heterotic NLSM be well-defined.
So, the conditions for the two A twists to be well-defined are
equivalent, just as for the anomaly-cancellation conditions.

Let us also check that the chiral ring of the Landau-Ginzburg
theory matches of the NLSM in the same universality class.
Recall from section~\ref{ssec:topTwists} that the relevant part of the chiral
ring in the (0,2) A twist is the hypercohomology of the complex
\begin{displaymath}
\cdots \: \Lambda^2 {\cal E}^{\vee} \: \stackrel{ i_{E^a} }{\longrightarrow}
\: {\cal E}^{\vee} \: \stackrel{ i_{E^a} }{\longrightarrow} \: {\cal O}_X.
\end{displaymath}
It can be shown that
the hypercohomology of this sequence,
for the ${\cal E}$ given above and $X$ the total space of the bundle,
is the same
\footnote{
We would like to thank T.~Pantev for providing an argument for this result,
outlined in appendix~\ref{app:hyper}.}
as
\begin{displaymath}
\oplus_i H^{*-1}\left(B, \Lambda^i {\cal E}'^{\vee} \right),
\end{displaymath}
so long as the $E^a$ are injective, precisely matching the corresponding
part of the chiral ring of the (0,2) A-twisted heterotic NLSM.
Similar results are true in models equally amenable to the B twist,
as we shall discuss in section~\ref{hetb}.

\subsection{General analysis of B twist}   \label{genl02ptwist}

We may also consider a (0,2) B twist of this
Landau-Ginzburg theory.  
Since the $E^a$ are nonzero, we cannot perform a na\"ive B twist,
but rather must twist the bosonic $p$ fibers of $X$ viewed as a vector bundle.

First, let us discuss anomaly cancellation.
There is a natural ${\bf C}^{\times}$ action on $X$, namely that which
multiplies the fibers of the vector bundle by phases, but leaves the base
$B$ invariant. This is the same one used in the twist of the $p$ fields.  
In order to make sense of anomalies, we need to, loosely speaking,
distinguish parts of the gauge bundle that live over the base $B$ from
those parts that live over the fibers $p$.
To do that, we shall use the following fact.
Given any vector bundle ${\cal E}$ on $X$ that is
equivariant with respect to the ${\bf C}^{\times}$ action on $X$,
its restriction to the zero section of $X$ (viewed as a vector bundle)
naturally splits
into a direct sum of components indexed by the characters $\chi$ of 
${\bf C}^{\times}$:
\begin{displaymath}
{\cal E}|_0 \: = \: \oplus_{\chi} {\cal E}_{\chi}.
\end{displaymath}
In the present case, ${\cal E} = \pi^* {\cal F}_2$, and so the restriction to
the zero section is invariant under the ${\bf C}^{\times}$:
${\cal E}_0 = \pi^* {\cal F}_2|_0$, ${\cal E}_{\chi} = 0$ for 
$\chi \neq 0$.  The tangent bundle of $X$, on the other hand, decomposes
nontrivially.  The result is that the anomaly cancellation condition for
the (0,2) B twist (involving the $p$ field) in this Landau-Ginzburg
theory to make sense is
that
\begin{displaymath}
 \Lambda^{top} {\cal F}_2  \: \cong \:
K_B  \otimes  \Lambda^{top} {\cal F}_1  \: ( \not\cong \: K_X|_0),
\end{displaymath} 
which precisely matches the condition for the (0,2) B twist of the heterotic
NLSM on $B$ to make sense:
\begin{displaymath}
K_B \: \cong \: \Lambda^{top} {\cal E}' \: \cong \:
\Lambda^{top} {\cal F}_2 \otimes \Lambda^{top}{\cal F}_1^{\vee},
\end{displaymath}
as expected, given that the theories are related by renormalization group
flow.

\subsection{Spectators}

Before going on, we should mention one other general point that might
concern the reader.  In the original
analysis of (0,2) GLSMs and Landau-Ginzburg models
over vector spaces in \cite{dk}, `spectator' fields were introduced
so as to insure that the Fayet-Iliopolous parameter did not flow
under the action of the renormalization group.
The analogous problem in the present case is that our space $X$
over which the Landau-Ginzburg  model is defined is not Calabi-Yau,
even if the space $B$ on which the corresponding IR NLSM lives is.  That could be fixed by modifying $X$ to be given by
\begin{displaymath}
\mbox{Tot} \left(
{\cal F}_1 \oplus \left( K_B \otimes \Lambda^{top} {\cal F}_1^{\vee}
\right) \: \stackrel{ \pi }{\longrightarrow} \: B \right),
\end{displaymath}
and the gauge bundle ${\cal E}$ to
\begin{displaymath}
\pi^* {\cal F}_2 \oplus
\pi^*
\left( K_B \otimes \Lambda^{top} {\cal F}_1^{\vee}
\right)^{\vee}.
\end{displaymath}
We then add a (0,2) superpotential term (an $F_a$) coupling
the extra direction $q$ of $X$ to the extra line bundle
factor in ${\cal E}$ -- the simplest form is $q \Omega$ with $\Omega$ the
Fermi superfield corresponding to the $\pi^*
\left( K_B \otimes \Lambda^{top} {\cal F}_1^{\vee}
\right)^{\vee}$ factor.  Then, $X$ is Calabi-Yau, but
the extra coordinate on $X$ pairs up with the extra factor in ${\cal E}$, which
after integrating them both out leaves us with the Landau-Ginzburg model we
have discussed so far.
Note furthermore that because the Chern character is a ring
homomorphism (meaning, $\mbox{ch}({\cal E} \oplus {\cal F})  = 
\mbox{ch}({\cal E}) + \mbox{ch}({\cal F})$, among other things),
the anomaly cancellation condition $\mbox{ch}_2(TX) = \mbox{ch}_2({\cal E})$
is satisfied in this Landau-Ginzburg theory if and only if it was
satisfied in the original one.  

Since we added
a nonzero $F_a$, the na\"ive (0,2)-analogue of the $A$ and $B$ twists no longer
make sense, though we can perform both if we twist the
bosonic fields.
For example, we can make sense of the (0,2) A twist if we twist the
$q$ field.  The consistency condition for the A twist to make sense
then is
\begin{displaymath}
\Lambda^{top} {\cal E}^{\vee} \: \cong \: 
K_{X_0} \otimes \pi_0^* \left( K_B \otimes 
\Lambda^{top} {\cal F}_1^{\vee} \right)^{\vee},
\end{displaymath}
where 
\begin{displaymath}
X_0 \: = \: \mbox{Tot}\left( {\cal F}_1 \: \stackrel{\pi_0}{\longrightarrow}
B \right)
\end{displaymath}
and
\begin{displaymath}
\Lambda^{top} {\cal E}^{\vee} \: \cong \:
\pi_0^* \Lambda^{top} {\cal F}_2^{\vee} \otimes
\pi_0^*\left( K_B \otimes \Lambda^{top} {\cal F}_1^{\vee} \right)^{\vee}.
\end{displaymath}
Thus, the consistency condition for the A-twisted theory (with spectators)
to make sense
is equivalent to
the condition
\begin{displaymath}
\pi_0^* \Lambda^{top} {\cal F}_2^{\vee} \: \cong \:
K_{X_0},
\end{displaymath}
which is the same as the consistency
condition for the original A-twisted theory,
without spectators.  So, adding spectators does not change
the condition for the A twist to be well-defined.
Similarly, we can perform the (0,2) B twist by twisting the $p$ field,
as in the last section.  The consistency condition for this twist,
in the theory with spectators, is
\begin{displaymath}
\Lambda^{top} {\cal F}_2 \otimes \left(
K_B \otimes \Lambda^{top} {\cal F}_1 \right)^{\vee}
\: \cong \: K_B \otimes \Lambda^{top} {\cal F}_1
\otimes \left( K_B \otimes \Lambda^{top} {\cal F}_1^{\vee} \right)^{\vee},
\end{displaymath}
which is equivalent to
\begin{displaymath}
\Lambda^{top} {\cal F}_2 
\: \cong \: K_B \otimes \Lambda^{top} {\cal F}_1.
\end{displaymath}
Note that this is precisely the condition for the B-twisted theory without spectators
to be consistent.
So, here again, adding spectators does not change the consistency condition
for the B twist.

Analogous constructions are possible in the other
heterotic examples discussed in this paper.  As spectators will not
impact anomalies or other computations, we will not discuss
them further.

\subsection{Example:  ${\bf P}^1 \times {\bf P}^1$}
\label{heta}

To be more specific,  
we shall consider a theory explored in
\cite{abs,ks1,ksb,ilarion1,ilarion2}:  a (0,2) theory on ${\bf P}^1 \times
{\bf P}^1$ with gauge bundle given by a deformation of the tangent bundle.
Specifically, we will consider bundles ${\cal E}'$ on 
${\bf P}^1 \times {\bf P}^1$ defined as cokernels of injective maps
\begin{equation}   \label{p1p1ex}
0 \: \longrightarrow \: {\cal O} \oplus {\cal O} \stackrel{ *
}{\longrightarrow}
{\cal O}(1,0)^2 \oplus {\cal O}(0,1)^2
\: \longrightarrow \: {\cal E}' \: \longrightarrow \: 0
\end{equation}
with
\begin{displaymath}
* \: = \:
\left[ \begin{array}{cc}
       x_1 & \epsilon_1 x_1  \\
       x_2 & \epsilon_2 x_2 \\
        0 & \widetilde{x_1} \\
        0 & \widetilde{x_2} 
\end{array} \right],
\end{displaymath}
where $x_1, x_2$ are homogeneous coordinates on one ${\bf P}^1$ and
$\widetilde{x_1}, \widetilde{x_2}$ are homogeneous coordinates on the other.
In the special case that $\epsilon_1 = \epsilon_2$,
the cokernel above becomes the tangent bundle to
${\bf P}^1 \times {\bf P}^1$.

To describe this theory, we will consider a Landau-Ginzburg model
on 
\begin{displaymath}
X \: = \: \mbox{Tot} \left( {\cal O} \oplus {\cal O}
\: \stackrel{\pi}{\longrightarrow} \: {\bf P}^1 \times {\bf P}^1 \right)
\end{displaymath}
with holomorphic vector bundle
${\cal E} = \pi^*{\cal O}(1,0)^2 \oplus \pi^* {\cal O}(0,1)^2$,
all $F_a \equiv 0$.
The $E^a$ are defined as follows.
Let $p_1$, $p_2$ denote the coordinates along the fibers of
${\cal O} \oplus {\cal O}$ in $X$, then the $E^a(\phi)$ are given by
\begin{eqnarray*}
E^1 & = & x_1 p_1 \: + \: \epsilon_1 x_1 p_2 \\
E^2 & = & x_2 p_1 \: + \: \epsilon_2 x_2 p_2 \\
E^3 & = & \widetilde{x_1} p_2 \\
E^4 & = & \widetilde{x_2} p_2.
\end{eqnarray*}
Note that the Landau-Ginzburg theory is defined over a NLSM instead of a
GLSM, so before substituting the equations above into physics one must
replace homogeneous coordinates by affine coordinates.

This Landau-Ginzburg model should renormalization-group flow to the
heterotic NLSM described above.
For later use, write
\begin{displaymath}
\left( \tilde{E}_1^a \right) \: = \: 
\left[ \begin{array}{c}
x_1 \\ x_2 \\ 0 \\ 0
\end{array} \right], \: \: \:
\left( \tilde{E}_2^a \right) \: = \:
\left[ \begin{array}{c}
\epsilon_1 x_1 \\ \epsilon_2 x_2 \\ \widetilde{x_1} \\
\widetilde{x_2} 
\end{array} \right],
\end{displaymath}
so that $E^a = p_1 \tilde{E}_1^a + p_2 \tilde{E}_2^a$.
Then, the superpotential terms have the form
\begin{displaymath}
2 \sum_a \left| p_1 \tilde{E}_1^a \: + \:
p_2 \tilde{E}_2^a \right|^2 \: + \:
\psi_+^i \lambda_-^{\overline{a}} \left(
p_1 D_i \tilde{E}_1^a \: + \:
p_2 D_i \tilde{E}_2^a \right) \: + \:
\psi_+^{p1} \lambda_-^{\overline{a}} \tilde{E}_a^a \: + \:
\psi_+^{p2} \lambda_-^{\overline{a}} \tilde{E}_2^a
\: + \: \mbox{cc},
\end{displaymath}
where $i$ denotes an index along ${\bf P}^1 \times {\bf P}^1$,
and $\psi_{\pm}^{p1}$, $\psi_{\pm}^{p2}$ are the superpartners 
of $p_1$, $p_2$.
Since $\tilde{E}_1^a$ and $\tilde{E}_2^a$ are nowhere-zero,
the first term above acts as a mass term for $p_1$ and $p_2$.
Furthermore, two linear combinations of the $\lambda_-$'s also
get a mass, namely the combinations
\begin{displaymath}
\lambda_-^{\overline{a}} \: = \: \lambda \overline{
\tilde{E}_1^a }, \:
 \lambda' \overline{
\tilde{E}_2^a }.
\end{displaymath}
For example, when we plug the first into the Yukawa coupling
\begin{displaymath}
i \psi_+^{p1} \lambda_-^{\overline{a}} \tilde{E}_a^a,
\end{displaymath}
we see that it becomes
\begin{displaymath}
i \psi_+^{p1} \lambda \left| \tilde{E}_1^a \right|^2,
\end{displaymath}
which acts as a mass term for $\lambda$ and $\psi_+^{p1}$.
(This is the usual mechanism by which the fermionic gauge symmetry
implicit in the $E^a$'s is realized physically, see for example
\cite{distrev}.)
Thus, the $p$ fields get a mass, as do some of the
$\lambda$'s, leaving us with the right data to describe a theory
that is surely in the same universality class as
the heterotic NLSM on ${\bf P}^1 \times {\bf P}^1$
with gauge bundle given by the deformation of the tangent bundle above.

Since the $F_a$ are all identically zero, we can define the
A analogue twist easily, without twisting any bosonic fields.
This means that all of the fields are twisted
in the form
\begin{align*}
\psi_+^i \: (\equiv \chi^i) \: &\in \: \Gamma_{C^{\infty}}\left( \phi^* T^{1,0}X \right)                                                                                   & \lambda_-^a \: (\equiv \lambda_{\overline{z}}^a) \: &\in \: \Gamma_{ C^{\infty} }\left( \overline{K}_{\Sigma} \otimes \left( \phi^* \overline{ {\cal E} } \right)^{\vee} \right) \\
\psi_+^{\overline{\imath}} \: (\equiv \psi_z^{\overline{\imath}}) \: &\in \: \Gamma_{ C^{\infty} }\left( K_{\Sigma} \otimes \left( \phi^* T^{1,0} X \right)^{\vee} \right) & \lambda_-^{\overline{a}} \: ( \equiv \lambda^{\overline{a}}) \: &\in \: \Gamma_{ C^{\infty} }\left( \phi^* \overline{ {\cal E} } \right).
\end{align*}

The superpotential interactions in this theory have the form
\begin{eqnarray*}
\lefteqn{
2 \sum_a \left| p_1 \tilde{E}_1^a \: + \:
p_2 \tilde{E}_2^a \right|^2 \: + \:
\chi^i \lambda^{\overline{a}} \left(
p_1 D_i \tilde{E}_1^a \: + \:
p_2 D_i \tilde{E}_2^a \right) \: + \:
\chi^{p1} \lambda^{\overline{a}} \tilde{E}_1^a \: + \:
\chi^{p2} \lambda^{\overline{a}} \tilde{E}_2^a
} \\
& \hspace*{0.5in} &
\: + \: 
\psi_z^{\overline{\imath}} \lambda^a_{\overline{z}} \left(
\overline{p}_1 D_{\overline{\imath}} \overline{ \tilde{E}_1^a } \: + \:
\overline{p}_2 D_{ \overline{\imath}} \overline{ \tilde{E}_2^a } \right)
\: + \:
\psi_z^{p1} \lambda_{\overline{z}}^a \overline{ \tilde{E}_1^a } \: + \:
\psi_z^{p2} \lambda_{\overline{z}}^a \overline{ \tilde{E}_2^a }.
\end{eqnarray*}

If we restrict to degree zero maps on a genus zero worldsheet
and also restrict to the contribution from zero modes
(as was done in \cite{ks1,bchir}), then we find that
\begin{eqnarray*}
\lefteqn{
\langle {\cal O}_1 \cdots {\cal O}_n \rangle  \: = \:
\int_{ {\bf P}^1 \times {\bf P}^1} d^2 x^i
\int_{ {\bf C}^2 } d^2 p_1 d^2 p_2
\int d \chi^i d \chi^{p1} d \chi^{p2}
\int d \lambda^{\overline{a}} \,
{\cal O}_1 \cdots {\cal O}_n 
} \\
& & \hspace*{1.5in} \cdot \exp\left(
- 2 \sum_a \left| p_1 \tilde{E}_1^a \: + \:
p_2 \tilde{E}_2^a \right|^2 \: - \:
\chi^i \lambda^{\overline{a}} \left(
p_1 D_i \tilde{E}_1^a \: + \:
p_2 D_i \tilde{E}_2^a \right) \right. \\
& & \hspace*{1.9in} \left. \: - \:
\chi^{p1} \lambda^{\overline{a}} \tilde{E}_1^a \: - \:
\chi^{p2} \lambda^{\overline{a}} \tilde{E}_2^a
\right).
\end{eqnarray*}

When we integrate out the $\chi^p$'s, this becomes
\begin{eqnarray*}
\lefteqn{
\langle  {\cal O}_1 \cdots {\cal O}_n \rangle \: = \:
\int_{ {\bf P}^1 \times {\bf P}^1} d^2 x^i
\int_{ {\bf C}^2 } d^2 p_1 d^2 p_2
\int d \chi^i 
\int d \lambda^{\overline{a}} \,
{\cal O}_1 \cdots {\cal O}_n 
\left( \lambda^{\overline{a}} \tilde{E}_1^a \right)
\left(  \lambda^{\overline{b}} \tilde{E}_2^b \right) 
} \\
& & \hspace*{1.5in} \cdot \exp\left(
- 2 \sum_a \left| p_1 \tilde{E}_1^a \: + \:
p_2 \tilde{E}_2^a \right|^2 \: - \:
\chi^i \lambda^{\overline{a}} \left(
p_1 D_i \tilde{E}_1^a \: + \:
p_2 D_i \tilde{E}_2^a \right) \right).
\end{eqnarray*}

Integrating out the $p$'s will merely generate some function
$f$ of the $\tilde{E}$'s:
\begin{equation}    \label{p1p1d0}
\langle {\cal O}_1 \cdots {\cal O}_n \rangle  \: = \:
\int_{ {\bf P}^1 \times {\bf P}^1} d^2 x^i
\int d \chi^i 
\int d \lambda^{\overline{a}} \,
{\cal O}_1 \cdots {\cal O}_n 
\left( \lambda^{\overline{a}} \tilde{E}_1^a \right)
\left(  \lambda^{\overline{b}} \tilde{E}_2^b \right) 
f\left(\tilde{E}_1^a, \tilde{E}_2^a\right).
\end{equation}

The expression above will give the same results for classical
correlation functions as in \cite{ks1}.  To see this, first note that
is that it is computing correlation functions between sheaf
cohomology $H^{\cdot}({\bf P}^1\times {\bf P}^1, {\cal E}'^{\vee})$
where ${\cal E}'^{\vee}$ is the kernel of the short exact sequence
\begin{displaymath}
0 \: \longrightarrow \: {\cal E}'^{\vee} \: \longrightarrow
\: {\cal O}(-1,0)^2 \oplus {\cal O}(0,-1)^2 \:
\stackrel{ \tilde{E} }{\longrightarrow} \: {\cal O} \oplus {\cal O}
\: \longrightarrow \: 0
\end{displaymath}
obtained by dualizing the sequence~(\ref{p1p1ex}).
The correlators ${\cal O}_i$ are representatives valued in
${\cal O}(-1,0)^2 \oplus {\cal O}(0,-1)^2$, and the factors of
$\lambda^{\overline{a}} \tilde{E}^a_i$ force the projection onto the
kernel of the map defined by the $\tilde{E}$'s, giving us sheaf cohomology
valued in ${\cal E}'^{\vee}$.  

More formally, from the long exact sequence of bundles on 
${\bf P}^1\times {\bf P}^1$, 
\begin{displaymath}
\cdots \: \longrightarrow \: H^i\left(
{\cal E}'^{\vee}
\right) \: \longrightarrow \:
H^i\left( {\cal O}(-1,0)^2 \oplus {\cal O}(0,-1)^2
\right) \: \longrightarrow \:
H^i\left( {\cal O} \oplus {\cal O} \right)
\: \longrightarrow \: \cdots
\end{displaymath}
we see that the representatives of $H^i({\cal E}'^{\vee})$ in
sheaf cohomology valued in ${\cal O}(-1,0)^2 \oplus {\cal O}(0,-1)^2$
are exactly those which are in the kernel of the map to ${\cal O} 
\oplus {\cal O}$, so given representatives ${\cal O}_i$ valued
in ${\cal O}(-1,0)^2 \oplus {\cal O}(0,-1)^2$, the $\lambda^{\overline{a}}
\tilde{E}^a$ factors are precisely insuring that the ${\cal O}_i$
represent sheaf cohomology valued in ${\cal E}'^{\vee}$.

Let us also check this more concretely.  From \cite{ks1}[equ'n (20)],
the classical correlation functions are given by
\begin{displaymath}
\langle \tilde{X}^2 \rangle  \: = \: \langle 1 \rangle  \: = \: 0, \: \: \:
\langle X \tilde{X} \rangle  \: = \: 1, \: \: \:
\langle X^2  \rangle \: = \: \epsilon_1 \: - \: \epsilon_2
\end{displaymath}
where $X$, $\tilde{X}$ are representatives of $H^{\cdot}({\bf P}^1 \times
{\bf P}^1, {\cal E}'^{\vee})$.  
In expression~(\ref{p1p1d0}), given the forms of $\tilde{E}^a_1$,
$\tilde{E}^a_2$, it is trivial to see that if $\tilde{X}$ has only 
$\lambda^{\overline{1}}$, $\lambda^{\overline{2}}$ factors,
and $X$ has only $\lambda^{\overline{3}}$, $\lambda^{\overline{4}}$
factors, then, $\langle \tilde{X}^2\rangle = 0$, $\langle X \tilde{X}\rangle \neq 0$,
and $\langle X^2\rangle \propto \epsilon_1 - \epsilon_2$, as expected.

It is straightforward to check that correlation functions in
nonperturbative sectors also match those in \cite{ks1}.
In a sector describing maps of degree $\vec{d} = (d_1,d_2)$,
it is straightforward to show that
\begin{eqnarray*}
\lefteqn{
\langle  {\cal O}_1 \cdots {\cal O}_n \rangle _{ \vec{d} } \: = \:
\int_{ {\cal M}_{\vec{d}} } d^2 x^i
\int_{ {\bf C}^2 } d^2 p_1 d^2 p_2
\int d \chi^i d \chi^{p1} d \chi^{p2}
\int d \lambda^{\overline{a}} \,
{\cal O}_1 \cdots {\cal O}_n 
} \\
& & \hspace*{1.5in} \cdot \exp\left(
- 2 \sum_a \left| p_1 \tilde{E}_1^a \: + \:
p_2 \tilde{E}_2^a \right|^2 \: - \:
\chi^i \lambda^{\overline{a}} \left(
p_1 D_i \tilde{E}_1^a \: + \:
p_2 D_i \tilde{E}_2^a \right) \right. \\
& & \hspace*{1.9in} \left. \: - \:
\chi^{p1} \lambda^{\overline{a}} \tilde{E}_1^a \: - \:
\chi^{p2} \lambda^{\overline{a}} \tilde{E}_2^a
\right),
\end{eqnarray*}
where ${\cal M}_{\vec{d}} = {\bf P}^{2d_1 + 1} \times {\bf P}^{2d_2 + 1}$
is the compactified moduli space of maps defined by the GLSM, as in
\cite{daveronen},
and the bundle ${\cal E}'$ induces \cite{ks1}
${\cal F} \rightarrow {\cal M}_{\vec{d}}$
given as the cokernel
\begin{displaymath}
0 \: \longrightarrow \: {\cal O} \oplus {\cal O} \: \stackrel{ \tilde{E} }{
\longrightarrow} \:
{\cal O}(1,0)^{2(d_1 + 1)} \oplus {\cal O}(0,1)^{2(d_2 + 1)} 
\: \longrightarrow \:
{\cal F} \: \longrightarrow \: 0,
\end{displaymath}
or, more pertinently,
\begin{displaymath}
0 \: \longrightarrow \: {\cal F}^{\vee} \: \longrightarrow \:
{\cal O}(-1,0)^{2(d_1 + 1)} \oplus {\cal O}(0,-1)^{2(d_2 + 1)} \: 
\stackrel{ \tilde{E} }{\longrightarrow} \: {\cal O} \oplus {\cal O} \: 
\longrightarrow \: 0.
\end{displaymath}
In the expression above, we have implicitly extended the
induced bundle over the compactification of the moduli space,
using the methods of \cite{ks1}.
Following the same analysis as before, we can write
\begin{displaymath}    
\langle {\cal O}_1 \cdots {\cal O}_n \rangle_{\vec{d}} \: = \:
\int_{ {\cal M}_{\vec{d}} } d^2 x^i
\int d \chi^i 
\int d \lambda^{\overline{a}} \,
{\cal O}_1 \cdots {\cal O}_n 
\left( \lambda^{\overline{a}} \tilde{E}_1^a \right)
\left(  \lambda^{\overline{b}} \tilde{E}_2^b \right) 
f'\left(\tilde{E}_1^a, \tilde{E}_2^a\right).
\end{displaymath}
Here, the ${\cal O}_i$ represent sheaf cohomology valued in 
\begin{displaymath}
{\cal O}(-1,0)^{2(d_1 + 1)} \oplus {\cal O}(0,-1)^{2(d_2 + 1)},
\end{displaymath}
and the factors of $\lambda^{\overline{a}} \tilde{E}^a$ in the
correlation function enforce a reduction to the kernel, giving
${\cal F}^{\vee}$.
So, the correlation function computes products of sheaf cohomology
on ${\cal M}_d$ valued in ${\cal F}^{\vee}$, exactly as needed
to match the results of \cite{ks1}.

\subsection{Example:  the quintic hypersurface in ${\bf P}^4$}
\label{ex:quinticdef}

Suppose we wish to describe in (0,2) language a (2,2) Landau-Ginzburg
model that flows in the IR to a NLSM on a complete
intersection in a space $B$ defined by $\{ G = 0 \} \subset B$, where $G$ is
a smooth section of some vector bundle ${\cal F} \rightarrow B$.
To do this, we could use the (0,2) version of (2,2) Landau-Ginzburg
models discussed earlier in this paper:
consider a (0,2) Landau-Ginzburg model over a space
\begin{displaymath}
X \: = \: \mbox{Tot }\left( {\cal F}^{\vee} \: \stackrel{ \pi }{
\longrightarrow }
\: B \right)
\end{displaymath}
with gauge bundle ${\cal E} = TX$,
$E^a \equiv 0$, and $F_a = \partial_a( p G)$ where $p$ is a fiber
coordinate on ${\cal F}^{\vee}$.
Anomaly cancellation is automatic, since ${\cal E} = TX$.
Judging from examples in \cite{dk,witphases} of which this is a straightforward
generalization\footnote{
For completeness, we outline the pertinent standard observations here.
The bosonic potential terms are of the form
$|G|^2 + |p|^2 \sum_i |D_i G|^2$.  So long as $G$ is smooth section,
it cannot be the case that $G=0$ and $D_i G=0$ simultaneously, so the
only way the potential can vanish is if $G = 0$ and $p=0$, so in the IR
the theory should flow to a NLSM on 
$\{ G=0 \} \cap \{p=0\} \subset X$.  Yukawa couplings enforce analogous
constraints on the fermionic superpartners. 
},
this theory should flow in the IR to a (2,2) NLSM
on $\{ G = 0 \} \subset B$.  
The fact that the B-twisted chiral ring of the heterotic
Landau-Ginzburg theory should match that of the NLSM
follows from a closely related discussion in section~\ref{sec:review}
and an analogous section in \cite{gs1}.

Given a Landau-Ginzburg model of the form above, we could modify it
to describe a heterotic NLSM with gauge bundle
a deformation of the tangent bundle, as follows.
Replace the
\begin{displaymath}
(F_a) \: = \: \partial_a \left( p G \right) \: = \:
\left( G, p D_i G \right)
\end{displaymath}
with
\begin{displaymath}
(F_a) \: = \: \left( G, p \left( D_i G \: + \: G_i \right) \right),
\end{displaymath}
but leave everything else the same -- for example, leave
${\cal E} = TX$.  This should flow in the IR to a heterotic NLSM with a gauge bundle different from the tangent bundle.

Let us now turn to a more specific example of a heterotic
Landau-Ginzburg model.
Let $X$ be the total space of the line bundle ${\cal O}(-5) \rightarrow
{\bf P}^4$, and consider a theory with (0,2) superpotential
\begin{displaymath}
\Lambda^p G(\phi) \: + \: \Lambda^i p \left( D_i G(\phi) \: + \: G_i
\right),
\end{displaymath}
where $G(\phi)$ is a section of ${\cal O}(5)$, $p$ a local coordinate
on the fiber of ${\cal O}(-5)$. The $\Lambda$'s are in one-to-one
correspondence with the chiral superfields, so that
they are defined by ${\cal E} = T^{1,0}X$, 
and all the $E^a \equiv 0$.
This model is very nearly the same as the (2,2) Landau-Ginzburg model
that is in the same universality class as the quintic,
discussed earlier, except for the fact that we are deforming
the tangent bundle via the $G_i(\phi)$ to a more general bundle.
In other words, we expect that this theory should renormalization-group flow
to a heterotic NLSM whose gauge bundle is a deformation
of the tangent bundle of the quintic in ${\bf P}^4$, 
defined by the cohomology of the short complex
\begin{displaymath}
0 \: \longrightarrow \: {\cal O}_Q \:
\stackrel{x_i}{\longrightarrow} \:
{\cal O}(1)^5|_Q \: \stackrel{ D_i G + G_i }{\longrightarrow}
\: {\cal O}(5)|_Q \: \longrightarrow \: 0,
\end{displaymath}
with $Q$ the quintic, defined by $\{ G = 0 \}$.

Now, let us perform the A twist analogue for this model.
Since the $F_a$ are not all zero, we must twist bosonic fields,
as described previously.  
Mechanically, this means that the fermions related to coordinates
on ${\bf P}^4$ couple to the bundles
\begin{align*}
\psi_+^i \: (\equiv \chi^i) \: &\in \: \Gamma_{C^{\infty}}\left( \phi^* T^{1,0}X \right)                                                                                   & \lambda_-^i \: (\equiv \lambda_{\overline{z}}^i) \: &\in \: \Gamma_{ C^{\infty} }\left( \overline{K}_{\Sigma} \otimes \left( \phi^* T^{0,1}X \right)^{\vee} \right) \\
\psi_+^{\overline{\imath}} \: (\equiv \psi_z^{\overline{\imath}}) \: &\in \: \Gamma_{ C^{\infty} }\left( K_{\Sigma} \otimes \left( \phi^* T^{1,0} X \right)^{\vee} \right) & \lambda_-^{\overline{\imath}} \: ( \equiv \lambda^{\overline{\imath}}) \: &\in \: \Gamma_{ C^{\infty} }\left( \phi^* T^{0,1}X \right).
\end{align*}
On the other hand, the $p$ field and its various fermionic partners
are twisted differently:
\begin{eqnarray*}
p \: \left(\equiv p_z \right) & \in & 
\Gamma_{ C^{\infty} }\left(K_{\Sigma} \otimes \phi^* T^{1,0}_{\pi} \right) \\
\overline{p} \: \left( \equiv \overline{p}_{ \overline{z} } \right)
& \in &
\Gamma_{ C^{\infty} }\left( \overline{K}_{\Sigma} \otimes
\phi^* T^{0,1}_{\pi} \right) 
\end{eqnarray*}
\begin{align*}
\psi_+^p \: (\equiv \psi_z^p) \: &\in \: \Gamma_{C^{\infty}}\left( K_{\Sigma} \otimes \phi^* T^{1,0}_{\pi} \right)                                & \lambda_-^p \: (\equiv \lambda^p) \: &\in \: \Gamma_{ C^{\infty} }\left( \left( \phi^* T^{0,1}_{\pi} \right)^{\vee} \right) \\
\psi_+^{\overline{p}} \: (\equiv \chi^{ \overline{p} }) \: &\in \: \Gamma_{ C^{\infty} }\left( \left( \phi^* T^{1,0}_{\pi} \right)^{\vee} \right) & \lambda_-^{\overline{p}} \: (\equiv \lambda_{\overline{z}}^{\overline{p}}) \: &\in \: \Gamma_{ C^{\infty} }\left( \overline{K}_{\Sigma} \otimes \phi^* T^{0,1}_{\pi} \right)
\end{align*}

Next, let us work out the correlators.
In the A analogue twist, the $\alpha_-$ are the scalar BRST parameters,
so we find that the BRST transformations of the fields are
\begin{align*}
		\delta \phi^i                      & =     i \alpha_- \chi^i                            & \delta p_z                                   & =     i \alpha_- \psi_z^p                             \\ 
		\delta \phi^{\overline{\imath}}    & =     0                                            & \delta \overline{p}_{\overline{z}}           & =     0                                               \\ 
		\delta \chi^i                      & =     0                                            & \delta \psi_z^p                              & =     0                                               \\ 
		\delta \psi_z^{\overline{\imath}}  & =     - \alpha_- \partial \phi^{\overline{\imath}} & \delta \chi^{\overline{p}}                   & =     - \alpha_- \partial \overline{p}_{\overline{z}} \\ 
		\delta \lambda^{\overline{\imath}} & =     0                                            & \delta \lambda^p                             & \neq  0                                               \\ 
		\delta \lambda_{\overline{z}}^i    & \neq  0                                            & \delta \lambda_{\overline{z}}^{\overline{p}} & =     0.     
\end{align*}
As a result, dimension-zero BRST-closed operators must be built from
$\phi^{\overline{\imath}}$, $\chi^i$, $\lambda^{\overline{\imath}}$.
On the face of it, the BRST cohomology looks na\"ively as if it is
 hypercohomology on $B$,
of the complex
\begin{displaymath}
\cdots \: \longrightarrow \: \Lambda^2 TB \: \longrightarrow \: TB \: \longrightarrow \:
{\cal O}_B
\end{displaymath}
using the fact that
\begin{displaymath}
\delta \left( h_{\overline{a} b} \lambda_-^b \right) \: = \:
i \alpha_- \overline{F}_{\overline{a}}
\end{displaymath}
to define the differential in the complex.  However, there are two
important subtleties.  First, along $p=0$, all of the $F_a$ appearing in
the complex above vanish, so we would get ordinary sheaf cohomology on $B$.
More importantly, the remarks concerning the chiral ring of the quintic in
section~3.3 of \cite{gs1} apply equally well here, so we must restrict to
the quintic hypersurface.

The superpotential terms in this theory have the form
\begin{align*}
 2 |G|^2  +  2 p_z \overline{p}_{\overline{z}} | D_i G + G_i |^2  
& + \chi^i \lambda_{\overline{z}}^j p_z D_i \left( D_j G + G_j \right) 
+  \psi_z^p \lambda_{\overline{z}}^j \left( D_j G + G_j \right)  +  \chi^i \lambda^p D_i G \\
& +  \psi_z^{\overline{\imath}} \lambda^{\overline{\jmath}} \overline{p}_{\overline{z}} D_{\overline{\imath}} \left( D_{\overline{\jmath}} \overline{G} 
+ \overline{G}_{ \overline{\jmath}} \right) 
+  \psi_z^{\overline{\imath}} \lambda_{\overline{z}}^{\overline{p}} D_{\overline{\imath}} \overline{G}  
+  \chi^{\overline{p}} \lambda^{\overline{\imath}} \left( D_{ \overline{\imath}} \overline{G} 
+ \overline{G}_{\overline{\imath}} \right).
\end{align*}

Therefore, if we restrict to degree zero maps on a genus zero
worldsheet and to the contribution
from zero modes (as was done in \cite{ks1,ksa,bchir,ksb}),
then we find that
\begin{eqnarray*}
\lefteqn{
\langle {\cal O}_1 \cdots {\cal O}_n \rangle \: = \:
\int d^2 \phi^i \int d \chi^i \int d \lambda^{\overline{\imath}}
\int d \chi^{\overline{p}} \int d \lambda^p \, 
{\cal O}_1 \cdots {\cal O}_n } \\
& & \hspace*{1in}\cdot \exp\left( \,
- \: 2 |G|^2 \: - \: \chi^i \lambda^p D_i G
\: - \: \chi^{\overline{p}} \lambda^{\overline{\imath}}
\left( D_{\overline{\imath}} \overline{G} + 
\overline{G}_{\overline{\imath}} \right) 
\: - \:
R_{i \overline{p} p \overline{k}} \chi^i
\chi^{\overline{p}} \lambda^p \lambda^{\overline{k}}
\right).
\end{eqnarray*}
The analogous (2,2) result \eqref{eq:quinticMQ} comprised a 
Mathai-Quillen form, as discussed in detail in \cite{gs1}.  
Here, however, we have a deformation
of a Mathai-Quillen form -- one only has a Mathai-Quillen form on the
(2,2) locus, and in particular only there does one expect convenient properties
such as scaling independence of the superpotential.

Integrating out $\chi^{\overline{p}}$ and $\lambda^p$ and omitting an
irrelevant factor of 2, this becomes
\begin{eqnarray*}
\lefteqn{
\langle {\cal O}_1 \cdots {\cal O}_n \rangle \: = \:
\int d^2 \phi^i \int d \chi^i \int d \lambda^{\overline{\imath}}
 \, 
{\cal O}_1 \cdots {\cal O}_n 
\left[
\left( \chi^i D_i G \right)
\left( \lambda^{\overline{\imath}} \left( D_{\overline{\imath}}
\overline{G} + \overline{G}_{\overline{\imath}} \right) \right)
\: + \:
R_{i \overline{p} p \overline{k}} g^{p \overline{p}} \chi^i \lambda^{
\overline{k}}
\right] 
} \\
& & \hspace*{5in} \cdot
\exp\left( - |G|^2 \right).
\end{eqnarray*}
In the scaling limit $\lambda \rightarrow 0$, where
$(D_i G + G_i) \mapsto \lambda (D_i G + G_i)$,
this reduces to
\begin{equation}
\langle {\cal O}_1 \cdots {\cal O}_n \rangle \: = \:
\int d^2 \phi^i \int d \chi^i \int d \lambda^{\overline{\imath}}
 \, 
{\cal O}_1 \cdots {\cal O}_n 
\left(
R_{i \overline{p} p \overline{k}} g^{p \overline{p}} \chi^i \lambda^{
\overline{k}}
\right),
\end{equation}
which matches the $\lambda \rightarrow 0$ scaling limit of the (2,2)
correlation function given in
\eqref{eq:quinticLimit0}.
In the scaling limit $\lambda \rightarrow \infty$, the correlation function
becomes
\begin{equation}
\langle {\cal O}_1 \cdots {\cal O}_n \rangle \: = \:
\int d^2 \phi^i \int d \chi^i \int d \lambda^{\overline{\imath}}
 \, 
{\cal O}_1 \cdots {\cal O}_n 
\left( \chi^i D_i G \right)
\left( \lambda^{\overline{\imath}} \left( D_{\overline{\imath}}
\overline{G} + \overline{G}_{\overline{\imath}} \right) \right)
\exp\left( - |G|^2 \right),
\end{equation}
which gives a different result than the corresponding (2,2)
scaling limit 
\eqref{eq:quinticLimitInf}.
In particular, correlation functions in this limit should match
those of the corresponding (0,2) NLSM:  just as in our previous discussion
of ${\bf P}^1 \times {\bf P}^1$ with a deformation of the tangent
bundle, the $\chi^i D_i G$ terms enforce a restriction to the
quintic hypersurface and the $\lambda^{\overline{\imath}}( D_{\overline{
\imath}} \overline{G} + \overline{G}_{\overline{\imath}})$
terms enforce a corresponding restriction on the gauge degrees of freedom.
 
The reason that the two scaling limits are giving different results
in this (0,2) theory, unlike the (2,2) theories discussed earlier,
is that the superpotential is no longer BRST-exact,
and so the twisted theory is not invariant under rescalings of the
superpotential.

One of many open problems in (0,2) quantum cohomology 
\cite{ks1,ade,ksa,bchir,ksb,ilarion1,ilarion2}
is that at present, no (0,2) analogue of a virtual
fundamental class computation is known.  Given that in the (2,2) theories
discussed earlier, very simple examples of virtual fundamental class
computations appeared in the scaling limit $\lambda \rightarrow \infty$ of
the superpotential, suggests that the $\lambda \rightarrow \infty$ limit
of the computation above should be a prototype for a (0,2) virtual fundamental
class computation.  We shall not pursue this matter further here, but it
is certainly worthy of further investigation.

\section{Models realizing kernels of maps}
\label{hetb}

Suppose we have a heterotic NLSM on a space $B$
with gauge bundle given by the kernel ${\cal E}'$ of the short
exact sequence
\begin{displaymath}
0 \: \longrightarrow \: {\cal E}' \: \longrightarrow \:
{\cal F}_1 \: \stackrel{F_a}{\longrightarrow} \: {\cal F}_2 \: 
\longrightarrow \: 0.
\end{displaymath}
Applying ideas from \cite{dk},
this heterotic NLSM should be in the same
universality class as a heterotic Landau-Ginzburg model 
on
\begin{displaymath}
X \: = \: \mbox{Tot}\left( {\cal F}_2^{\vee} \: \stackrel{\pi}{\longrightarrow}
\: B \right),
\end{displaymath}
with gauge bundle ${\cal E} = \pi^* {\cal F}_1$, $E^a \equiv 0$,
and $F_a = p \tilde{F}_a$ 
defined by the map $\tilde{F}_a: {\cal F}_1 \rightarrow {\cal F}_2$
defining ${\cal E}'$ and $p$ fiber coordinates on ${\cal F}_2$.

The arguments are standard, but for completeness, let us briefly outline
why these two theories should be in the same universality class.
The bosonic potential term $h^{a \overline{b}} F_a \overline{F}_{\overline{b}}$
becomes $h^{a \overline{b}} |p|^2 \tilde{F}_a 
\overline{\tilde{F}}_{\overline{b}}$, which due to the surjectivity of $F$
acts as a mass term for $p$.  Thus, in the IR, the theory should flow 
to a theory on $B$ instead of $X$.  The Yukawa couplings include
$i \psi_+^p \lambda_-^a \tilde{F}_a$, which gives a mass to $\psi_+^p$
and those fermions $\lambda_-^a$ not in the kernel of $F$, so that
the gauge bundle in the IR should be ${\cal E}'$.
As further checks, we shall next argue that anomalies and chiral rings
of the two theories match.

Let us now show that anomaly cancellation conditions in the
Landau-Ginzburg and NLSMs match.
Anomaly cancellation in the NLSM on $B$ is the
statement that
\begin{displaymath}
\mbox{ch}_2(TB) \: = \: \mbox{ch}_2({\cal E}') \: = \:
\mbox{ch}_2({\cal F}_1) \: - \: \mbox{ch}_2({\cal F}_2).
\end{displaymath}
Anomaly cancellation in the heterotic Landau-Ginzburg model over $X$,
on the other hand, is the statement that
\begin{displaymath}
\mbox{ch}_2(TX) \: = \: \mbox{ch}_2({\cal E}) \: = \:
\pi^* \mbox{ch}_2({\cal F}_1).
\end{displaymath}
From the definition of $X$, there is a short exact
sequence
\begin{displaymath}
0 \: \longrightarrow \: \pi^* {\cal F}_2^{\vee} \: \longrightarrow \:
TX \: \longrightarrow \: \pi^* TB \: \longrightarrow \: 0,
\end{displaymath}
from which we read off that
\begin{displaymath}
\mbox{ch}_2(TX) \: = \: \pi^* \mbox{ch}_2(TB) \: + \:
\pi^* \mbox{ch}_2({\cal F}_2),
\end{displaymath}
and hence the anomaly-cancellation conditions in the NLSM and in the Landau-Ginzburg model are equivalent:
anomaly cancellation holds in one model if and only if it
holds in the other model.

Since the $E^a \equiv 0$ in these heterotic Landau-Ginzburg models,
one should be able to perform the B twist na\"ively, without twisting any
bosonic fields.  When one does so, the consistency condition in the
heterotic Landau-Ginzburg model for the twist to be sensible is that
\begin{displaymath}
\Lambda^{top} {\cal E} \: \cong \: K_X.
\end{displaymath}
Comparing to the heterotic NLSM,
\begin{displaymath}
K_X \: \cong \: \pi^*\left( K_B \otimes \Lambda^{top} {\cal F}_2 \right), \:
\: \:
\Lambda^{top} {\cal E}^{\vee} \: \cong \:
\pi^* \Lambda^{top} {\cal F}_1,
\end{displaymath}
so we see that the condition for the B twist of the heterotic Landau-Ginzburg
model to be consistent is just
\begin{displaymath}
\pi^* \Lambda^{top} {\cal F}_1 \: \cong \:
\pi^* K_B \otimes \pi^* \Lambda^{top} {\cal F}_2
\end{displaymath}
or equivalently
\begin{displaymath}
\pi^* K_B \: \cong \: \pi^* \Lambda^{top} {\cal F}_1 \otimes \pi^* \Lambda^{top}
{\cal F}_2^{\vee} \: \cong \: \pi^* \Lambda^{top} {\cal E}',
\end{displaymath}
which is the pullback of the constraint that the B twist of the corresponding
heterotic NLSM be well-defined.  So, the conditions that the
two B twists be well-defined are equivalent to one another, just as for
the anomaly-cancellation conditions.

Let us next check that chiral rings match.
Recall from section~\ref{hetlggenl} that the relevant part of  
the chiral ring in the (0,2) B-twisted heterotic Landau-Ginzburg model is given
by
the hypercohomology on $X$ of the sequence
\begin{displaymath}
\cdots \: \Lambda^2 {\cal E} \: \stackrel{ i_{F_a}}{\longrightarrow} \:
{\cal E} \: \stackrel{ i_{F_a} }{\longrightarrow} \:
{\cal O}_X
\end{displaymath}
(the $F_a$ assumed surjective)
which, using arguments closely analogous to those outlined in
section~\ref{heta} above,  
is the same as
\begin{displaymath}
\oplus_i H^{*-1}\left(B, \Lambda^i {\cal E}' \right),
\end{displaymath}
exactly right to match the corresponding part of the
chiral ring of the (0,2) B-twisted
heterotic NLSM which should be in the same universality class.

If we twist the $p$ fields, then we can also perform the (0,2) A twist
of this theory.  Proceeding as in section~\ref{genl02ptwist},  
we work on the zero section of $X$, where the consistency condition for
the (0,2) A twist of this Landau-Ginzburg model is
\begin{displaymath}
\lambda^{top} {\cal F}_1^{\vee} 
\: \cong \: K_B \otimes \Lambda^{top} {\cal F}_2^{\vee}
\: ( \not\cong K_X|_0).
\end{displaymath}
This is the same as
\begin{displaymath}
K_B \: \cong \: \Lambda^{top} {\cal F}_1^{\vee} \otimes \Lambda^{top} {\cal F}_2
\: \cong \: \Lambda^{top} ({\cal E}')^{\vee},
\end{displaymath}
which is exactly the condition for the (0,2) A twist of the IR NLSM to be well-defined, as expected.

In fact, the general observations here are precisely dual to
those in section~\ref{heta}, which should not be surprising in light
of the dualities we have discussed between heterotic A and B analogue twistings.
Formally, starting from the data above, we could obtain a pair of
theories (Landau-Ginzburg, NLSM) amenable to an A twist
rather than a B twist by replacing the surjective map
\begin{displaymath}
F_a: \: {\cal F}_1 \: \longrightarrow \: {\cal F}_2
\end{displaymath}
with the injective map
\begin{displaymath}
F_a^*: \: {\cal F}_2^{\vee} \: \longrightarrow \: {\cal F}_1^{\vee}
\end{displaymath}
(also replacing each occurrence of ${\cal F}_1$ by ${\cal F}_2^{\vee}$
and so forth),
and taking $E^a = F_a^*$, and the new $F_a$ to be identically zero
(matching the fact that the old $E^a$ vanished).

\section{Models realizing cohomologies of monads}  \label{sec:monad}

Suppose we have a heterotic NLSM on a space $B$
with gauge bundle given by the cohomology of the short complex
\begin{displaymath}
0 \: \longrightarrow \: {\cal F}_1 \: \stackrel{\tilde{E}^a}{\longrightarrow} \:
{\cal F}_2 \: \stackrel{\tilde{F}_a}{\longrightarrow} \: {\cal F}_3 \:
\longrightarrow \: 0
\end{displaymath}
at the middle term.
Judging from related examples and standard analyses in previous sections
here and in
\cite{dk}, this heterotic NLSM should be in the same
universality class as a heterotic Landau-Ginzburg model 
on
\begin{displaymath}
X \: = \: \mbox{Tot}\left( 
{\cal F}_1 \oplus {\cal F}_3^{\vee} \: \stackrel{\pi}{\longrightarrow} \:
B \right),
\end{displaymath}
with ${\cal E} \equiv \pi^*{\cal F}_2$,
and $E^a = p' \tilde{E}^a$, $F_a = p \tilde{F}_a$, 
where $p$ are fiber coordinates along ${\cal F}_3^{\vee}$ and $p'$ fiber
coordinates along ${\cal F}_1$.

This statement about universality classes can be justified in exactly
the same form as before.  For brevity, let us omit the completely
standard analysis of masses, and instead turn to checks that anomalies
match.
Specifically, let us
briefly check that the anomaly cancellation conditions in these
two models are equivalent.

In the heterotic Landau-Ginzburg model,
anomaly cancellation is the condition that
\begin{displaymath}
\mbox{ch}_2(TX) \: = \: \mbox{ch}_2({\cal E}) \: = \:
\mbox{ch}_2({\cal F}_2),
\end{displaymath}
where $\mbox{ch}_2(TX) = \pi^* \mbox{ch}_2(TB) + 
\pi^* \mbox{ch}_2({\cal F}_1) + \pi^* \mbox{ch}_2({\cal F}_3)$.
Thus, anomaly cancellation in the heterotic Landau-Ginzburg model can
be rewritten as
\begin{displaymath}
\pi^* \mbox{ch}_2(TB) \: = \:
\pi^* \mbox{ch}_2({\cal F}_2) \: - \:
\pi^* \mbox{ch}_2({\cal F}_1) \: - \:
\pi^* \mbox{ch}_2({\cal F}_3),
\end{displaymath}
which is just the pullback of the anomaly cancellation condition in the
heterotic NLSM.
Thus, anomaly cancellation is the same constraint in the two theories:
anomaly cancellation in one is equivalent to anomaly cancellation
in the other.

For reasons of brevity we do not include here any correlation function
computations in these most general heterotic Landau-Ginzburg models.

\section{Models realizing monads over complete intersections}
\label{sec:monadCI}

Here, we consider the most general case.
Suppose we want a heterotic Landau-Ginzburg model that will flow to
a heterotic NLSM on a complete intersection
$Y \equiv \{ G_{\mu} = 0 \} \subset B$ 
defined by $G_{\mu} \in \Gamma({\cal G})$,
${\cal G}$ a holomorphic vector bundle on $B$,
with a gauge bundle ${\cal E}'$ given by the cohomology of the short complex of 
holomorphic vector bundles
\begin{displaymath}
0 \: \longrightarrow \: {\cal F}_1|_Y \: \stackrel{\tilde{E}^a|_Y}{
 \longrightarrow} \:
{\cal F}_2|_Y \: \stackrel{\tilde{F}_a|_Y}{\longrightarrow} 
\: {\cal F}_3|_Y \: \longrightarrow \: 0,
\end{displaymath}
where $\tilde{E}_a: {\cal F}_1 \rightarrow {\cal F}_2$
and $\tilde{F}_a: {\cal F}_2 \rightarrow {\cal F}_3$ are defined over
all of $B$, but the sequence above only necessarily becomes a complex
over $Y \subset B$.
Generalizing the (0,2) GLSM description in \cite{dk},
the corresponding Landau-Ginzburg model is defined over the space
\begin{displaymath}
X \: = \: \mbox{Tot}\left( {\cal F}_1 \oplus
{\cal F}_3^{\vee} \: \stackrel{\pi}{\longrightarrow} \: B
\right),
\end{displaymath}
with gauge bundle ${\cal E}$ an extension\footnote{
In general, the extension will be nontrivial, as an example we will
discuss momentarily will make clear.  Aside from that, we have not 
found a way to uniquely determine the extension in terms of
data of the IR NLSM.
In fact, since renormalization group flow is a lossy process, it is
not completely clear that the Landau-Ginzburg model should be
uniquely determined by the NLSM -- perhaps several
different extensions defining different ${\cal E}$'s in the Landau-Ginzburg
model all flow to the same NLSM.  We have no such examples,
but neither can we rule out the possibility.
} of $\pi^* {\cal F}_2$ by
$\pi^* {\cal G}^{\vee}$:
\begin{displaymath}
0 \: \longrightarrow \: \pi^* {\cal G}^{\vee} \: \longrightarrow \:
{\cal E} \: \longrightarrow \: \pi^* {\cal F}_2 \: \longrightarrow
\: 0.
\end{displaymath}
The $F_a \in \Gamma({\cal E}^{\vee})$ are partially determined by
$G \in \Gamma({\cal G})$ and
\begin{displaymath}
F_a |_{ \pi^* {\cal F}_2^{\vee} } \: = \: p \tilde{F}_a,
\end{displaymath}
where $p$ are fiber coordinates on ${\cal F}_3^{\vee}$
and $\tilde{F}_a$ is the map ${\cal F}_2 \rightarrow {\cal F}_3$.
The $E^a \in \Gamma({\cal E})$ are partially determined by $p' \tilde{E}^a$,
where $p'$ are fiber coordinates on ${\cal F}_1$ and
$\tilde{E}^a$ is the map ${\cal F}_1 \rightarrow {\cal F}_2$.

In the special case that we wish to describe the cohomology of a monad
over $B$, and not a complete intersection in $B$, then we
take ${\cal G} \equiv 0$ and then 
the data above trivially reduces to that
discussed in the previous section.

Let us also see how the examples discussed in
section~\ref{ex:quinticdef} are a special case of this construction.
There, a deformation of a (2,2) model on a complete intersection was
considered, and the corresponding Landau-Ginzburg model on the (2,2)
locus was described by
\begin{displaymath}
X \: = \: \mbox{Tot }\left( {\cal G}^{\vee} \: \stackrel{\pi}{
\longrightarrow } B \right),
\end{displaymath}
with $E^a = 0$, $F_a = \partial_a (p G)$, $G$ a section of ${\cal G}$,
and gauge bundle ${\cal E} = TX$.
We recover this as a special case of the current construction by
taking ${\cal F}_1=0$, ${\cal F}_3 = {\cal G}$, and ${\cal F}_2 = TB$,
and utilizing the short exact sequence
\begin{displaymath}
0 \: \longrightarrow \: \pi^* {\cal G}^{\vee} \: \longrightarrow
\: TX \: \longrightarrow \: \pi^* TB \: \longrightarrow \: 0.
\end{displaymath}
The tangent bundle to the complete intersection $Y =
\{ G_{\mu} = 0 \} \subset B$
is the kernel of the map
\begin{displaymath}
TB|_Y \: \stackrel{ D G_{\mu} }{ \longrightarrow} \: {\cal G}|_Y
\end{displaymath}
and hence is the cohomology of the monad, that defines the gauge
bundle over the NLSM.

Finally, the gauge bundle in the Landau-Ginzburg model
is given by the extension
\begin{displaymath}
0 \: \longrightarrow \: \pi^* {\cal G}^{\vee} \: \longrightarrow \:
{\cal E} \: \longrightarrow \: \pi^* TB \: \longrightarrow \: 0,
\end{displaymath}
which also defines $TX$, hence, ${\cal E} = TX$, exactly right to match
the (2,2) theory discussed in section~\ref{ex:quinticdef}.

Let us now check how anomaly cancellation in the Landau-Ginzburg model
compares to anomaly-cancellation in the corresponding NLSM.  In the Landau-Ginzburg model, anomaly cancellation is the statement
that
\begin{displaymath}
\mbox{ch}_2(TX) \: = \: \mbox{ch}_2({\cal E}),
\end{displaymath}
but here 
\begin{displaymath}
\mbox{ch}_2({\cal E}) \: = \:
\pi^* \mbox{ch}_2({\cal G}) \: + \:
\pi^* \mbox{ch}_2({\cal F}_2),
\end{displaymath}
and similarly
\begin{displaymath}
\mbox{ch}_2(TX) \: = \: \pi^* \mbox{ch}_2(TB) \: + \: 
\pi^* \mbox{ch}_2({\cal F}_1) \: + \: \pi^* \mbox{ch}_2({\cal F}_3).
\end{displaymath}
So we see that $\mbox{ch}_2({\cal E}) = \mbox{ch}_2(TX)$ implies that
\begin{displaymath}
\pi^* \mbox{ch}_2(TB) \: - \: \pi^* \mbox{ch}_2({\cal G})\: = \:
\pi^* \mbox{ch}_2({\cal F}_2) \: - \: \pi^* \mbox{ch}_2({\cal F}_1)
\: - \: \pi^* \mbox{ch}_2({\cal F}_3),
\end{displaymath}
but the right-hand side of the expression above is the second Chern character
of the pullback of the gauge bundle in the NLSM,
and the left-hand side is the second Chern character of the pullback of the
tangent bundle to the complete intersection, so we see that anomaly
cancellation in the Landau-Ginzburg model is the pullback of the anomaly
cancellation condition in the NLSM to which it flows,
as expected.

By twisting bosonic fields, we can perform the (0,2) A and B twists.
Let us compare the consistency conditions for those twists in the 
Landau-Ginzburg model to the consistency conditions for the (0,2) A and
B twists of the IR NLSM.  Since they are related by
renormalization group flow, the consistency conditions should match.

First, we shall consider the (0,2) A twist.
Here we twist the $p$ field, that gives local coordinates on the
fibers of ${\cal F}_3^{\vee}$, as well as the fermionic fields in the
${\cal G}$ part of ${\cal E}$.
We define
\begin{displaymath}
X_0 \: = \: \mbox{Tot}\left( {\cal F}_1 \: \stackrel{ \pi_0 }{\longrightarrow}
\: B \right).
\end{displaymath}
Then, working on $X_0$, which is a zero section of $X$, the consistency condition
for the (0,2) A twist of the Landau-Ginzburg model is
\begin{displaymath}
\pi_0^* \left(  \Lambda^{top} {\cal F}_2^{\vee}
\right) \otimes \pi_0^*\left( \Lambda^{top} {\cal G}^{\vee} \right)
 \: \cong \:
K_{X_0} \otimes \pi_0^* \Lambda^{top} {\cal F}_3^{\vee}
\: ( \not\cong K_X |_0).
\end{displaymath}
Using the fact that 
\begin{displaymath}
K_{X_0} \: \cong \: \pi_0^* K_B \otimes \pi_0^* {\cal F}_1^{\vee},
\end{displaymath}
we see that the consistency condition for the (0,2) A twist of the
Landau-Ginzburg model is
\begin{displaymath}
\pi_0^* K_B \otimes \pi_0^* \Lambda^{top} {\cal G}\: \cong \: 
\pi_0^* \left( \Lambda^{top} {\cal F}_2^{\vee} \otimes \Lambda^{top} 
{\cal F}_3 \otimes \Lambda^{top} {\cal F}_1 \right).
\end{displaymath}
On the other hand, the consistency condition for the (0,2) A twist
of the NLSM (which does not require twisting any
bosonic fields) is given by
\begin{displaymath}
K_Y \: \cong \: \Lambda^{top} ({\cal E}')^{\vee} \: \cong \:
\Lambda^{top} {\cal F}_2^{\vee}|_Y \otimes
\Lambda^{top} {\cal F}_1|_Y \otimes
\Lambda^{top} {\cal F}_3|_Y.
\end{displaymath}
Using the fact that 
\begin{displaymath}
K_Y \: \cong \: i^* K_B \otimes i^* \Lambda^{top} {\cal G},
\end{displaymath}
where $i: Y \hookrightarrow B$ is inclusion, we see that the
consistency condition for the (0,2) A twist of the Landau-Ginzburg
model is the pullback along $i_0: Y \hookrightarrow X_0$ ($i =
\pi_0 \circ i_0$)
of the consistency condition for the (0,2) A twist of the IR NLSM, 
exactly as required by the renormalization group. 

The (0,2) B twist operates in a closely parallel fashion.
Here, we twist the $p'$ field (that gives local coordinates on the fibers
of ${\cal F}_1$), instead of $p$.  Define
\begin{displaymath}
X_1 \: = \: \mbox{Tot} \left( {\cal F}_3^{\vee} \: \stackrel{\pi_1}{
\longrightarrow }
\: B \right).
\end{displaymath}
Working on $X_1$, a zero section of $X$, the consistency condition for the
(0,2) B twist of the Landau-Ginzburg model is
\begin{displaymath}
\pi_1^*\left( \Lambda^{top} {\cal G}^{\vee} \otimes
\Lambda^{top} {\cal F}_2 \right) \: \cong \:
K_{X_1} \otimes \pi_1^* \Lambda^{top} {\cal F}_1 \:
( \not\cong K_X |_0).
\end{displaymath}
Using the fact that
\begin{displaymath}
K_{X_1} \: \cong \: \pi_1^* K_B \otimes \pi_1^* {\cal F}_3 
\end{displaymath}
we see that the consistency condition for the (0,2) B twist of
the Landau-Ginzburg model is
\begin{displaymath}
\pi_1^* K_B \otimes \pi_1^* {\cal G} \: \cong \:
\pi_1^* \left( \Lambda^{top} {\cal F}_2 \otimes
\Lambda^{top} {\cal F}_1^{\vee} \otimes
\Lambda^{top} {\cal F}_3^{\vee} \right).
\end{displaymath}
On the other hand, the consistency condition for the (0,2) B twist of
the NLSM (which does not require twisting any bosonic
fields) is given by
\begin{displaymath}
K_Y \: \cong \: \Lambda^{top} {\cal E}' 
\: \cong \:
\Lambda^{top} {\cal F}_2|_Y \otimes
\Lambda^{top} {\cal F}_1^{\vee}|_Y \otimes
\Lambda^{top} {\cal F}_3^{\vee}|_Y.
\end{displaymath}
Using the fact that
\begin{displaymath}
K_Y \: \cong \: i^* K_B \otimes i^* \Lambda^{top} {\cal G},
\end{displaymath}
we see that the consistency condition for the (0,2) B twist of the
Landau-Ginzburg model is the pullback of the consistency condition for
the (0,2) B twist of the IR NLSM, exactly as required by the renormalization
group.

\section{Conclusions}

In this paper, we have discussed heterotic Landau-Ginzburg models and 
analogues of topological twists therein, as well as checked our results by comparing
Landau-Ginzburg models and NLSMs in the same universality class.  

A future direction that should be pursued involves the heterotic
generalization of virtual fundamental class computations.  We have seen in
\cite{gs1} how ordinary Landau-Ginzburg computations give a physical
understanding of some simple special cases of virtual fundamental class
constructions, and we outlined in this work how heterotic Landau-Ginzburg
models compute correlation functions in a closely analogous fashion.
Therefore, presumably encoded within those heterotic Landau-Ginzburg model
computations are at least some simple cases of the heterotic generalization
of virtual fundamental classes.  In previous work \cite{ks1}, it was
outlined how one could in principle deal with the analogue of multicovers,
and how Euler classes of induced bundles were realized in sheaf cohomology
via Atiyah classes.  Heterotic Landau-Ginzburg models seem to also be
describing analogous effects via working on total spaces of normal bundles
and intersecting with zero sections.

Another interesting heterotic generalization would be to couple heterotic
Landau-Ginzburg models to the fibered WZW models of \cite{distshar}.

\appendix
\section{A hypercohomology computation}    \label{app:hyper}

In this appendix, we consider an A-twisted heterotic Landau-Ginzburg model
flowing to a NLSM on a manifold $B$, with a bundle ${\cal E}'$ defined as the
cokernel of an injective map:
\begin{displaymath}
{\cal E}' \: = \: \mbox{coker }\left\{
{\cal F}_1 \: \stackrel{\tilde{E}}{\longrightarrow} \: {\cal F}_2 \right\}.
\end{displaymath}
The Landau-Ginzburg model is defined over the space 
\begin{displaymath}
X \: = \: \mbox{Tot}\left( {\cal F}_1 \: \stackrel{\pi}{\longrightarrow}
\: B \right),
\end{displaymath}
with gauge bundle ${\cal E} \equiv \pi^* {\cal F}_2$,
all $F_a \equiv 0$,
and $E^a = p \tilde{E}^a$ for
$p$ fiber coordinates along ${\cal F}_1$.
In particular, we outline here an argument \cite{tonypriv622}  that
the hypercohomology of the complex
\begin{displaymath}
\cdots \: \longrightarrow \: \Lambda^2 {\cal E}^{\vee} \: \stackrel{ i_{E^a} }{\longrightarrow}
\: {\cal E}^{\vee} \: \stackrel{ i_{E^a} }{\longrightarrow} \: {\cal O}_X
\end{displaymath}
(which defines the theory's chiral ring) is given by 
\begin{displaymath}
\oplus_i H^{*-1}\left(B, \Lambda^i {\cal E}'^{\vee} \right)
\end{displaymath}
whenever the $E^a$ are injective, precisely matching the corresponding
part of the chiral ring of the (0,2) A-twisted heterotic NLSM.

The map $\pi : \mbox{Tot}({\cal F}_{1}) \to B$ is affine and so the  
pushforward functor is exact.  In particular, the hypercohomology of the 
complex $(\Lambda^{*}\pi^{*}{\cal E}^{\vee},i_{E^a})$ on 
$\mbox{Tot}({\cal F}_{1})$ 
is the              
same as the hypercohomology of the complex 
$(\pi_{*}\pi^{*}\Lambda^{*}{\cal E}^{\vee},i_{E^a})$ on $B$.  
On the other hand we have          
$\pi_{*}\pi^{*}\Lambda^{*}{\cal E}^{\vee} = 
\Lambda^{*}{\cal E}^{\vee}\otimes            
S^{*}{\cal F}_1^{\vee}$, where $S$ denotes symmetrized products. 
This is a consequence of the projection formula \cite{hart}[exercise II.5.1],
that for
any sheaf ${\cal S}$ on $B$ we have
$\pi_* \pi^* {\cal S} = {\cal S} \otimes \pi_* {\cal O} =
{\cal S} \otimes S^* {\cal F}_1^{\vee}$.

Thus, we may equivalently compute the hypercohomology of the complex                
\begin{displaymath} 
\ldots \wedge^{2}{\cal E}^{\vee}\otimes S^{*}{\cal F}_1^{\vee} 
\to {\cal E}^{\vee}\otimes S^{*}{\cal F}_1^{\vee}  
\to S^{*}{\cal F}_1^{\vee}                                                   
\end{displaymath} 
on $B$.  Note that the homological degree in this complex is the degree of
the wedge powers of ${\cal E}^{\vee}$, whereas the degree in the symmetric
powers of ${\cal F}_1^{\vee}$ plays no homological role.  On the other hand
$S^{*}{\cal F}_1^{\vee}$ is a direct sum of pieces of given homogeneous
degree, so  the whole complex decomposes in to a direct sum of complexes.
The first is the degree zero piece ${\cal O}_{B}$, the second is the
complex ${\cal E}^{\vee} \to {\cal F}_1^{\vee}$, the third is the complex
$\Lambda^{2}{\cal E}^{\vee} \to {\cal E}^{\vee}\otimes {\cal F}_1^{\vee}
\to S^{2} {\cal F}_1^{\vee}$, and so on until we reach the rank $r$ of
${\cal E}'$.  The $k^\text{th}$ complex is   
\begin{displaymath}                                                          
\Lambda^{k}{\cal E}^{\vee} \to \Lambda^{k-1}{\cal E}^{\vee}\otimes 
{\cal F}_1^{\vee} 
\to \cdots  \to
{\cal E}^{\vee}\otimes S^{k-1}{\cal F}_1^{\vee} \to S^{k}{\cal F}_1^{\vee}.
\end{displaymath}
For any $k > r$, the complexes are exact and so do not contribute           
to the cohomology.                                                              
                                                                         
On the other hand, each of the above complexes is exact in all terms             
except the first one, where the kernel is $\Lambda^{i}{\cal E}'^{\vee}$. 
So for              
$0 \leq i \leq r$ we have that the $i$-th complex is quasi-isomorphic to 
$\Lambda^{i}{\cal E}'^{\vee}$ placed in degree $(-i)$, 
and for $i > r$ we get that the  
$i$-th complex is exact.  
                                                                                
This shows that the hypercohomology is                         
$\oplus_{i} H^{*-i}(B,\wedge^{i}{\cal E}'^{\vee})$.

\section{Acknowledgments}

We would like to thank P.~Clarke, R.~Donagi, 
S.~Katz, E.~Witten, 
and especially I.~Melnikov and T.~Pantev for useful conversations.
In particular, we would like to thank I.~Melnikov for initial collaboration
and many useful conversations, and T.~Pantev for providing the
hypercohomology computations used in this paper.
J.G. was partially supported by NSF grant DMS-02-44412.
E.S. was partially supported by NSF grant DMS-0705381.


\begin{thebibliography}{199}

\addcontentsline{toc}{section}{References}

\bibitem{abs} A. Adams, A. Basu, S. Sethi, ``(0,2) duality,''
Adv. Theor. Math. Phys. {\bf 7} (2004) 865-950,
{\tt hep-th/0309226}.

\bibitem{ks1} S. Katz, E. Sharpe, ``Notes on certain (0,2)
correlation functions,'' Comm. Math. Phys. {\bf 262} (2006) 611-644,
{\tt hep-th/0406226}.

\bibitem{ade} A. Adams, J. Distler, M. Ernebjerg,
``Topological heterotic rings,''
Adv. Theor. Math. Phys. {\bf 10} (2006) 657-682,
{\tt hep-th/0506263}.

\bibitem{ksa} E. Sharpe, ``Notes on correlation functions in
(0,2) theories,'' {\tt hep-th/0502064}.

\bibitem{bchir} E. Sharpe, ``Notes on certain other (0,2) correlation
functions,'' {\tt hep-th/0605005}.

\bibitem{distshar} J. Distler, E. Sharpe, ``Heterotic compactifications
with principal bundles for general groups and general levels,''
{\tt  hep-th/0701244}.

\bibitem{mason-skinner} L. Mason, D. Skinner, ``Heterotic twistor-string
theory,'' {\tt arXiv:  0708.2276}.

\bibitem{ksb} J. Guffin, S. Katz, ``Deformed quantum cohomology and
(0,2) mirror symmetry,'' {\tt arXiv:  0710.2354}.

\bibitem{ilarion1} I. Melnikov, S. Sethi, ``Half-twisted (0,2) Landau-Ginzburg
models,'' {\tt arXiv:  0712.1058}.

\bibitem{ilarion2} J. McOrist, I. Melnikov, ``Half-twisted correlators from
the Coulomb branch,'' {\tt arXiv:  0712.3272}.

\bibitem{gs1} J. Guffin, E. Sharpe, ``A-twisted Landau-Ginzburg models,''
{\tt arXiv: 0801.3836}.

\bibitem{tonypriv622} T. Pantev, private communication, June 22, 2007.

\bibitem{dg} J. Distler, B. Greene, ``Aspects of (2,0) string
compactifications,'' Nucl. Phys. {\bf B304} (1988) 1-62.

\bibitem{dk} J. Distler, S. Kachru, ``(0,2) Landau-Ginzburg theory,''
Nucl. Phys. {\bf B413} (1994) 213-243,
{\tt hep-th/9309110}.

\bibitem{distrev} J. Distler, ``Notes on (0,2) superconformal field
theories,'' {\tt hep-th/9502012}.

\bibitem{daveronen} D. Morrison and M.R. Plesser,
``Summing the instantons:  quantum cohomology and mirror symmetry
in toric varieties,'' Nucl. Phys. {\bf B440} (1995) 279-354,
{\tt hep-th/9412236}.

\bibitem{witphases} E. Witten, ``Phases of ${\cal N}=2$ theories
in two dimensions,'' Nucl. Phys. {\bf B403} (1993) 159-222,
{\tt hep-th/9301042}.

\bibitem{hart} R. Hartshorne, {\it Algebraic geometry},
Springer-Verlag, New York, 1977.



\end{thebibliography}
\end{document}